\documentclass[a4paper,fleqn,usenatbib]{mnras}
\usepackage{longtable}
\usepackage{newtxtext,newtxmath}
\usepackage[T1]{fontenc}
\usepackage{ae,aecompl}
\usepackage{graphicx}	
\defcitealias{ccm89}{CCM89}
\defcitealias{ngc5904}{Paper I}
\defcitealias{ngc6205}{Paper II}
\defcitealias{ngc288}{Paper III}
\defcitealias{stetson2019}{SPZ19}
\defcitealias{zloczewski2012}{ZKR12}
\defcitealias{narloch2017}{NKP17}
\defcitealias{bonatto2013}{BCK13}
\defcitealias{vasiliev2021}{VB21}
\defcitealias{hendricks2012}{HSV12}
\defcitealias{sfd98}{SFD98}

\title[Isochrone fitting of NGC\,6362 and NGC\,6723]
{Isochrone fitting of Galactic globular clusters -- IV. NGC\,6362 and NGC\,6723}

\author[G. A. Gontcharov et al.]{
George~A.~Gontcharov,$^{1}$\thanks{E-mail: georgegontcharov@yahoo.com}
Maxim~Yu.~Khovritchev,$^{1,2}$
Aleksandr~V.~Mosenkov,$^{3,1}$
\newauthor
Vladimir~B.~Il'in,$^{1,2,4}$
Alexander~A.~Marchuk,$^{1,2}$
Denis~M.~Poliakov,$^{1,2}$
\newauthor
Olga~S.~Ryutina,$^{2}$
Sergey~S.~Savchenko,$^{1,2,5}$
Anton~A.~Smirnov,$^{1,2}$
Pavel~A.~Usachev,$^{1,2,5}$
\newauthor
Jae-Woo~Lee,$^{6}$
Conner~Camacho,$^{3}$
and Noah~Hebdon$^{3}$
\\
$^{1}$Central (Pulkovo) Astronomical Observatory, Russian Academy of Sciences, Pulkovskoye chaussee 65/1, St. Petersburg 196140, Russia\\
$^{2}$Saint Petersburg State University, Universitetskij pr. 28, St. Petersburg 198504, Russia\\
$^{3}$Department of Physics and Astronomy, N283 ESC, Brigham Young University, Provo, UT 84602, USA\\
$^{4}$Saint Petersburg University of Aerospace Instrumentation, Bol. Morskaya ul. 67A, St. Petersburg 190000, Russia\\
$^{5}$Special Astrophysical Observatory, Russian Academy of Sciences, 369167 Nizhnij Arkhyz, Russia\\
$^{6}$Department of Physics and Astronomy, Sejong University, 209 Neungdo-ro, Gwangjin-Gu, Seoul 05006, Korea\\
}

\date{Accepted 2022 November 08. Received 2022 October 26; in original form 2021 November 30}

\pubyear{2022}

\begin{document}
\label{firstpage}
\pagerange{\pageref{firstpage}--\pageref{lastpage}}
\maketitle

\begin{abstract}
We present new isochrone fits to the colour--magnitude diagrams of the Galactic globular clusters NGC\,6362 and NGC\,6723. We utilize 22 and 26 photometric filters for 
NGC\,6362 and NGC\,6723, respectively, from the ultraviolet to mid-infrared using data sets from {\it HST}, {\it Gaia}, unWISE, and other photometric sources.
We use models and isochrones from the Dartmouth Stellar Evolution Database (DSED) and Bag of Stellar Tracks and Isochrones (BaSTI) for $\alpha$--enhanced [$\alpha$/Fe]$=+0.4$ and 
different helium abundances.
The metallicities [Fe/H]$=-1.04\pm0.07$ and $-1.09\pm0.06$ are derived from the red giant branch slopes in our fitting for NGC\,6362 and NGC\,6723, respectively. 
They agree with spectroscopic estimates from the literature. We find a differential reddening up to $\Delta E(B-V)=0.13$ mag in the NGC\,6723 field due to the adjacent 
Corona Australis cloud complex. We derive the following for NGC\,6362 and NGC\,6723, respectively:
distances $7.75\pm0.03\pm0.15$ (statistic and systematic error) and $8.15\pm0.04\pm0.15$ kpc; ages $12.0\pm0.1\pm0.8$ and $12.4\pm0.1\pm0.8$ Gyr;
extinctions $A_\mathrm{V}=0.19\pm0.04\pm0.06$ and $0.24\pm0.03\pm0.06$ mag; reddenings $E(B-V)=0.056\pm0.01\pm0.02$ and $0.068\pm0.01\pm0.02$ mag.
DSED provides systematically lower [Fe/H] and higher reddenings than BaSTI. However, the models agree in their relative estimates:
NGC\,6723 is $0.44\pm0.04$ kpc further, $0.5\pm0.1$ Gyr older, $\Delta E(B-V)=0.007\pm0.002$ more reddened, and with $0.05\pm0.01$ dex lower [Fe/H] than NGC\,6362.
The lower metallicity and greater age of NGC\,6723 with respect to NGC\,6362 explain their horizontal branch morphology difference.
This confirms age as the second parameter for these clusters. We provide lists of the cluster members from the {\it Gaia} Data Release 3.
\end{abstract}


\section{Introduction}
\label{intro}

In recent years, the key ingredients for productive isochrone fitting of colour--magnitude diagrams (CMDs) of Galactic globular clusters (GCs) have been significantly improved. 
On the one hand, accurate photometry of individual stars in ultraviolet (UV), optical, and infrared (IR) bands has been obtained and presented in data sets
from the {\it Hubble Space Telescope (HST}; \citealt{piotto2015,nardiello2018,simioni2018}), {\it Gaia} Data Release 2 (DR2; \citealt{evans2018}) and
Early Data Release 3 (EDR3; \citealt{lindegren2021a,riello2021}), {\it Wide-field Infrared Survey Explorer (WISE}; \citealt{wise}) 
as the unWISE catalogue \citep{unwise}, various ground-based telescopes by \citet[][hereafter SPZ19]{stetson2019}, and other sources. 
On the other hand, theoretical stellar evolution models, such as
the Dartmouth Stellar Evolution Database (DSED, \citealt{dotter2007})\footnote{\url{http://stellar.dartmouth.edu/models/}} and
a Bag of Stellar Tracks and Isochrones (BaSTI, \citealt{pietrinferni2021})\footnote{\url{http://basti-iac.oa-abruzzo.inaf.it/index.html}},
have been upgraded for more accurate isochrones to predict CMDs with multiple low metallicity $\alpha$-enhanced stellar 
populations with primordial or enriched helium abundance, which are typical in GCs \citep{monelli2013,milone2017}.
Moreover, {\it Gaia} EDR3 provides very precise proper motions (PMs), which can be used for an accurate selection of GC members.

In \citet[][hereafter Paper I]{ngc5904}, \citet[][hereafter Paper II]{ngc6205}, and \citet[][hereafter Paper III]{ngc288} we fit CMDs for GCs NGC\,288, NGC\,362, NGC\,5904 (M5), 
NGC\,6205 (M13), and NGC\,6218 (M12) by various isochrones for different stages of stellar evolution.
They include the main sequence (MS), its turn-off (TO), the subgiant branch (SGB), red giant branch (RGB), horizontal branch (HB), and asymptotic giant branch (AGB).
This fitting allows us to derive age, distance, and reddening for a dominant population or a mix of populations in each CMD, except for some in the UV and IR.
Data set cross-identification allows us to estimate systematic differences between the data sets, convert the derived reddenings into extinctions in all the filters under 
consideration, and draw an empirical extinction law (i.e. a dependence of extinction on wavelength) for each combination of cluster, model and data set.

The pilot \citetalias{ngc5904} showed that such an approach is productive. 
In \citetalias{ngc6205}, we paid special attention to the balance of uncertainties in our approach
(see also Sect.~\ref{systematics}).
In \citetalias{ngc288}, we verified that a higher precision of the derived parameters can be achieved via an analysis of relative estimates for 
several similar GCs.

In this paper, we apply the same approach to the pair of GCs NGC\,6362 and NGC\,6723. This study is an enhancement of our approach due to the following: 
(i) these clusters are contaminated by foreground stars to a point where the data sets should be cleaned with PMs;
(ii) NGC\,6723 is affected by an adjacent foreground cloud complex;
(iii) the clusters are similar in metallicity, age, distance, and reddening, which allow us to verify the sensitivity of our approach in an analysis of 
the relative estimates;
(iv) the clusters are suitable for deriving [Fe/H] as a free parameter instead of adopting it from the literature.

As this is the fourth paper in this series, many details of our analysis, which we perform in this study, are given in our previous papers. 
We refer the reader to those papers, especially to the description of the balance of uncertainties, presented in appendix A of \citetalias{ngc6205},
creation of fiducial sequences (ridge lines), presented in section 3 of \citetalias{ngc6205}, adjustment of different data sets with adjacent filters, presented in
section 6 of \citetalias{ngc6205}, and usage of the {\it Gaia} EDR3 PMs and parallaxes for identification of cluster members, presented in section 3.2 of \citetalias{ngc288}.

Previous isochrone fittings were done by \citet{brocato1999,piotto1999,paust2010,kerber2018} for NGC\,6362, 
by \citet{alcaino1999} for NGC\,6723, and 
by \citet{dotter2010,vandenberg2013,omalley2017,wagner2017,valcin2020,oliveira2020} for both the clusters.
Their results can be compared with ours.

This paper is organized as follows. Some key properties of NGC\,6362 and NGC\,6723 are presented in Sect.~\ref{clusters}.
The theoretical models and corresponding isochrones, used for our isochrone-to-CMD fitting, as well as systematic uncertainties of the isochrones, are considered in Sect.~\ref{iso}.
In Sect.~\ref{photo}, we describe the initial data sets used, their cleaning, cluster member identification, and fiducial sequence creation in the CMDs.
In Sect.~\ref{results}, we present and discuss the results of our isochrone fitting. We summarize our main findings and conclusions in Sect.~\ref{conclusions}.
Some additional CMDs of the clusters are shown in Appendix~\ref{addcmds}.

\begin{table*}
\def\baselinestretch{1}\normalsize\normalsize
\caption[]{Some properties of the clusters under consideration. 
The {\it Gaia} EDR3 median parallax is calculated in Sect.~\ref{edr3}.
}
\label{properties}
\[
\begin{tabular}{lcc}
\hline
\noalign{\smallskip}
 Property            &  NGC\,6362  &  NGC\,6723 \\
\hline
\noalign{\smallskip}
R.A. J2000 (h~m~s) from \citet{goldsbury2010}                                 & \hphantom{$-$}17 31 55      &     \hphantom{$-$}18 59 33  \\
Decl. J2000 ($\degr$ $\arcmin$ $\arcsec$) from \citet{goldsbury2010}          & $-67$ 02 54                 &     $-36$ 37 56              \\
Galactic longitude ($\degr$) from \citet{goldsbury2010}                       & 325.55452                   &     0.06928                \\
Galactic latitude ($\degr$) from \citet{goldsbury2010}                        & $-17.56977$                 &     $-17.29893$              \\
Angular diameter (arcmin) from \citet{bica2019}                               &              19             &                 17           \\
Distance from the Sun (kpc) from \citet{harris}, 2010 revision\footnotemark\  & 7.6                         &     8.7                      \\
Distance from the Sun (kpc) from \citet{baumgardt2021}                        &      $7.65\pm0.07$          &     $8.27\pm0.10$            \\
{\it Gaia} EDR3 median parallax (mas) from this study                         &   $0.1230\pm0.011$           &   $0.1231\pm0.011$          \\ 
$[$Fe$/$H$]$ from \citet{carretta2009}                                        & $-1.07\pm0.05$              &  $-1.10\pm0.07$              \\
Mean differential reddening $\overline{\Delta E(B-V)}$ (mag) from \citetalias{bonatto2013}         & $0.025\pm0.008$   &   $0.027\pm0.009$  \\
Maximum differential reddening $\Delta E(B-V)_\mathrm{max}$ (mag) from \citetalias{bonatto2013}    & 0.046             &   0.051            \\
$E(B-V)$ (mag) from \citet{harris}, 2010 revision               & 0.09     & 0.05 \\
$E(B-V)$ (mag) from \citetalias{sfd98}                          & 0.07     & 0.16 \\ 
$E(B-V)$ (mag) from \citet{schlaflyfinkbeiner2011}              & 0.06     & 0.14 \\ 
$E(B-V)$ (mag) from \citet{planck}                              & 0.11     & 0.13 \\
\hline
\end{tabular}
\]
\end{table*}
\footnotetext{The commonly used database of GCs by \citet{harris} (\url{https://www.physics.mcmaster.ca/~harris/mwgc.dat}), 2010 revision.}

\section{Properties of the clusters}
\label{clusters}

Some general properties of NGC\,6362 and NGC\,6723 are presented in Table~\ref{properties}.

Each of the clusters has two populations \citep{dalessandro2014,mucciarelli2016,milone2017,lee2019}.
Both the populations of both the clusters are $\alpha$--enriched with $0.3<$[$\alpha$/Fe]$<0.4$ \citep{rojas2016,massari2017,crestani2019}.
\citet{milone2017} and \citet{oliveira2020} estimate the fraction of the first (primordial) population of stars in NGC\,6362 as 
$0.574\pm0.035$ and $0.584\pm0.041$, respectively, and $0.363\pm0.017$ and $0.377\pm0.029$ in NGC\,6723, respectively.
These estimates ensure that both the populations should be well represented in any CMD.
The populations differ in helium abundance by $\Delta Y\approx0.02$ or even $0.01$ \citep{vandenberg2018,milone2018,lagioia2018,lee2019}.
Therefore, being well represented in a CMD, the populations are nevertheless segregated in colour or magnitude only in some domains of some CMDs,
as discussed in Sect.~\ref{fiducials}. For an unresolved mix of the populations, which we meet in most CMD domains, we adopt the helium
abundance $Y\approx0.26$, i.e. 0.01 dex higher than $Y\approx0.25$ for the primordial population.

Table~\ref{properties} shows that NGC\,6362 and NGC\,6723 have moderately precise metallicity estimates.
A large diversity of [Fe/H] estimates from the literature for both the clusters can be found in \citet{lee2014,kaluzhny2015,arellano2018,vandenberg2018,vandenberg2022} and 
references therein:
$-1.26<$[Fe/H]$<-0.70$ for NGC\,6362 and $-1.35<$[Fe/H]$<-0.93$ for NGC\,6723.
Recent spectroscopic [Fe/H] estimates for NGC\,6723 show a large variety as well: $-1.22\pm0.08$ \citep{gratton2015}, $-0.98\pm0.08$ \citep{rojas2016}, 
$-0.93\pm0.05$ \citep{crestani2019}.
In contrast, recent spectroscopic [Fe/H] estimates for NGC\,6362 are consistent: $-1.09\pm0.01$ \citep{mucciarelli2016} and $-1.07\pm0.01$ \citep{massari2017}.

The slope of the RGB is sensitive to [Fe/H] 
\footnote{The same is also true for the faint MS, about $>4$ mag fainter than TO. However, isochrones have large systematic uncertainties in 
the faint MS domain. Hence, we do not consider this domain at all.}, 
since the continuum opacity of RGB stars is mainly caused by the H$^-$ ion with the metals being the major donor of free electrons. 
This allows us to derive [Fe/H] as an isochrone fitting parameter (together with distance, age, and reddening) in CMDs with well-populated RGB
in order to decrease fitting residuals. We then use our average [Fe/H] estimates for fitting the remaining CMDs. We do it separately for each model.

The RGBs are affected by saturation effects, crowding at the centres of the cluster fields, differential reddening, systematic errors of photometry, and helium enrichment 
(for the latter see \citealt{savino2018}). All these effects may lead to a typical systematic uncertainty of about 0.15 dex in our [Fe/H] estimate derived from CMD and a model.

Although both the clusters show an extended HB, where both the blue and red sides of the RR~Lyrae instability strip are populated, NGC\,6362 and NGC\,6723 are richer in red 
and blue HB stars, respectively.
Their HB morphology is defined by their metallicity, as the most important parameter, and also by a yet unrecognized second parameter (see \citealt{crestani2019}).
Thus, the similarity of NGC\,6362 and NGC\,6723 makes them an interesting, albeit poorly investigated, second-parameter pair.
In \citetalias{ngc288}, we have shown that age may be the second parameter for NGC\,288, NGC\,362, NGC5904, and NGC\,6218.
In this paper, we intend to answer whether age is the second parameter for NGC\,6362 and NGC\,6723 as well.

Table~\ref{properties} shows a rather low foreground and differential reddening\footnote{In this paper, we consider differential reddenings as systematic 
variations of colours in a cluster field due to variations of reddening, as well as due to other reasons, as discussed by \citet{anderson2008} and in 
\citetalias{ngc288}. For NGC\,6723, variations of reddening certainly dominate in systematic variations of colours in the cluster field.} 
for NGC\,6362 and NGC\,6723.
However, the differential reddening, presented in Table~\ref{properties}, is estimated by \citet[][hereafter BCK13]{bonatto2013} from {\it HST} ACS photometry within about 
3.4 arcmin of the cluster centres, i.e. only in a small central part of the cluster field.
Moreover, taking into account the stated precision of the reddening estimates in Table~\ref{properties} as a few units of the last decimal place, the inconsistency of these 
estimates is evident.

An obvious source of this inconsistency for NGC\,6723 is the adjacent Corona Australis cloud complex.
This complex contains dark and bright nebulae: IC\,4812, NGC\,6729 around Herbig Ae/Be star R~CrA, and a combination of NGC\,6726 and NGC\,6727 around Herbig Ae/Be star TY~CrA.
The latter is only 30 arcmin south (in Galactic coordinates) of NGC\,6723. Periphery of this cluster is obscured by periphery of the nebula. 
This is seen in some Digital Sky Survey images and star counts, as discussed in Sect.~\ref{edr3}.
The reddening maps of \citet[][hereafter SFD98]{sfd98} and \citet{planck} show $E(B-V)>1.7$ mag in the centre of NGC\,6726/NGC\,6727 nebula. 
In contrast, $E(B-V)\approx0.38$ and 0.11 mag on the edges of NGC\,6723, at points earest and farthest from the nebula, respectively (adopting the cluster diameter of 17 arcmin). 
In combination with the reddening estimates $E(B-V)\approx0.15$ mag from these maps for the NGC\,6723 centre (see Table~\ref{properties}), these maps suggest a high and non-linear 
differential reddening in the NGC\,6723 field.\footnote{For comparison, for NGC\,6362 the \citetalias{sfd98} map shows only $\Delta E(B-V)<0.01$ mag over the whole cluster field. 
Hence, we do not need a differential reddening correction for NGC\,6362.} Moreover, the brightness of the nebula parts is variable due to variability of TY~CrA.
This may lead to reddening estimate diversity in Table~\ref{properties} if the estimates are obtained in different parts of the cluster field or in different moments.
Surprisingly, to our knowledge, the Corona Australis cloud complex has never been considered as a source of differential reddening in the NGC\,6723 field. 
For example, \citet[][hereafter HSV12]{hendricks2012} described this cluster as `nearly unreddened GC NGC 6723'.

\section{Theoretical models and isochrones}
\label{iso}

To fit the CMDs of NGC\,6362 and NGC\,6723, we use the following theoretical models of stellar evolution and related $\alpha$--enhanced isochrones:
\begin{enumerate}
\item BaSTI \citep{newbasti,pietrinferni2021} with various [Fe/H], helium abundance $Y=0.25$ for primordial and $0.275$ for helium-enriched population,
[$\alpha$/Fe]$=+0.4$, initial solar $Z_{\sun}=0.0172$ and $Y_{\sun}=0.2695$, overshooting, diffusion, mass loss efficiency $\eta=0.3$, 
where $\eta$ is the free parameter in Reimers law \citep{reimers}.
We draw both the BaSTI isochrones with $Y=0.25$ and $0.275$ in all our CMD figures, while the interpolated isochrones with $Y=0.26$ are not drawn for clarity.
As in \citetalias{ngc288}, we also use the BaSTI extended set of zero-age horizontal branch (ZAHB) models with different values of the total mass but the same mass for the 
helium core and the same envelope chemical stratification. 
This set presents a realistic description of stochastic mass loss between the MS and HB, when stars with the same mass during the MS can lose different amounts of mass during 
the RGB and, hence, differ in their colours and magnitudes during the HB. This BaSTI set of ZAHB models appears very important for appropriate HB fitting.
\item DSED \citep{dotter2008} with various [Fe/H], helium abundance $Y=0.25$ for primordial and $0.33$ for helium-enriched population, [$\alpha$/Fe]$=+0.4$, solar $Z_{\sun}=0.0189$ 
and no mass loss. We use the $Y=0.25$ and $0.33$ isochrones in order to interpolate intermediate isochrones (DSED provides no $\alpha$--enhanced isochrone for $0.25<Y<0.33$). 
Naturally, the isochrones with $Y=0.25$ and $Y=0.26$ are close to each other, while $Y=0.33$ seems to overestimate the helium enrichment of these clusters (see Sect.~\ref{clusters}).
DSED gives no prediction for the HB and AGB. DSED does not provide isochrones for VISTA filters. However, we have verified that $J_\mathrm{VISTA}$ can be substituted by 
$J_\mathrm{UKIDSS}$, a filter from the United Kingdom Infrared Telescope Infrared Deep Sky Survey (UKIDSS; \citet{ukidss}), with a precision better than 0.01 mag.
We use the DSED isochrones for $J_\mathrm{UKIDSS}$ instead of $J_\mathrm{VISTA}$, while there is no DSED substitution for $Ks_\mathrm{VISTA}$ filter.
\end{enumerate}

DSED and BaSTI are models that provide user-friendly online tools to calculate $\alpha$--enhanced isochrones for various levels of helium enrichment and almost all filters under 
consideration, using stellar metallicity, age, and mass as input. Hereafter, we use only DSED and BaSTI $\alpha$--enhanced isochrones. 
This is in line with, for example, the isochrone fitting of the same clusters by \citet{oliveira2020}.

We consider the isochrones for a grid of some reasonable [Fe/H] with a step of 0.05 dex, distances with a step of 0.1 kpc, reddenings with a step of 0.001 mag, and ages over 
8 Gyr with a step of 0.5 Gyr.

In contrast to our previous studies, we derive [Fe/H] as an isochrone fitting parameter (together with distance, age, and reddening) in CMDs with a rich bright RGB 
(Table~\ref{cmds} shows the derived [Fe/H] estimates) and then use our average [Fe/H] estimates (separately for each model) for fitting the remaining CMDs.

The derived [Fe/H] are discussed and compared with those avaliable in the literature in Sect.~\ref{metal}.

\subsection{Systematics from isochrones}
\label{systematics}

Systematic errors from stellar modeling, bolometric corrections, and colour versus effective temperature (T$_\mathrm{eff}$) relations should be revealed in
(i) comparison of different isochrones in their fitting to similar data sets (model-to-model differences),
(ii) comparison of cluster parameter estimates from different studies (study-to-study differences),
(iii) theoretical evaluation of intrinsic uncertainty in each isochrone prediction ingredient and their unification.
The latter approach should be realized by the authors of models, while the others can be briefly realized here.

On the one hand, the only manifestation of the DSED and BaSTI model-to-model systematics in our final results is a systematically lower metallicity (by about [Fe/H]$=0.12$ dex) 
through DSED (see Sect.~\ref{metal}). The lower the metallicity the bluer the isochrone. Hence, by DSED, a lower [Fe/H] leads to systematically higher reddenings of about 
$\Delta E(B-V)=0.02$ mag.
On the other hand, a study-to-study scatter of recent spectroscopic [Fe/H] measurements shows $\pm0.14$ dex for NGC\,6723 and reads much lower for NGC\,6362 (see Sect.~\ref{properties}).
Therefore, assuming uncertainty of adopted [Fe/H] as the dominant contributor to uncertainty of reddening and extinction, we assign a systematic uncertainty corresponding 
to $\sigma E(B-V)=0.02$ or $\sigma A_\mathrm{V}=0.06$ to all our reddening/extinction results.

Both the model-to-model and the study-to-study comparison shows a rather large systematic uncertainty of age.
For example, a typical difference between age estimates, obtained for similar data sets by similar methods, from \citet{dotter2010,wagner2017,kerber2018,oliveira2020,valcin2020} is 
up to $\pm0.8$ Gyr.
\citet{vandenberg2018} noted that the age uncertainty is $\pm0.8$ Gyr, `of which $\pm0.5$ Gyr is due to a $\pm0.05$ mag uncertainty in the distance modulus and the rest 
corresponds to the net effect of metal abundance uncertainties'.
\citet{oliveira2020} noted `a conservative uncertainty of 0.5 Gyr in the ages can be adopted due to chemical abundance uncertainties'.
\citet{valcin2020} noted `In total, we have a 0.5 Gyr uncertainty budget due to systematic effects in stellar modeling'.
The difference of about 0.7 Gyr between the results of the single- and two-population analysis of \citet{wagner2016} and \citet{wagner2017} for the same clusters may be another 
manifestation of large systematic uncertainties. Finally, we assign 0.8 Gyr as a conservative systematic uncertainty of our derived ages.

All recent (since \citealt{vandenberg2013}) estimates of the distance modulus for NGC\,6362 using isochrone-to-CMD fitting, are within 14.36--14.45 (including our own 14.45 
from Sect.~\ref{results}). Those for NGC\,6723 are within 14.53--14.59. Taking into account the discussion in \citet{valcin2020} and the conclusion of \citet{vandenberg2022} 
`isochrones generally reproduce the main features of observed CMDs to within $\approx$0.03 mag', we adopt a systematic uncertainty of distance moduli as $\pm0.04$. 
This converts to $\pm150$ pc distance uncertainty for our clusters.

It is worth noting that if we adopt a systematic uncertainty from a study-to-study comparison, then, naturally, our result with this systematic uncertainty agrees with all 
the stuidies and, hence, {\it a posteriori} comparison of our result with those from the studies has little sense.

\section{Data sets}
\label{photo}

\subsection{Initial data sets}
\label{datasets}

We use the following data sets for both the clusters (hereafter {\it twin} data sets) -- see Table~\ref{filters}:
\begin{enumerate}
\item the {\it HST} Wide Field Camera 3 (WFC3) UV Legacy Survey of Galactic Globular Clusters (the $F275W$, $F336W$ and $F438W$ filters) and the
Wide Field Channel of the Advanced Camera for Surveys (ACS; the $F606W$ and $F814W$ filters) survey of Galactic globular clusters 
\citep{piotto2015,nardiello2018},\footnote{\url{http://groups.dfa.unipd.it/ESPG/treasury.php}}
\footnote{We do not use the {\it HST} ACS photometry from \citet{simioni2018} due to its sparsely populated RGB, SGB, and TO.}
\item photometry from \citet{piotto2002} in the $F439W$ and $F555W$ filters from the {\it HST} Wide Field and Planetary Camera 2 
(WFPC2),
%
\item $UBVRI$ photometry described by \citetalias{stetson2019}
\footnote{\url{http://cdsarc.u-strasbg.fr/viz-bin/cat/J/MNRAS/485/3042}}, 
with the NGC\,6362 and NGC\,6723 data sets processed within the same pipeline and presented 
recently,\footnote{\url{https://www.canfar.net/storage/vault/list/STETSON/homogeneous/Latest_photometry_for_targets_with_at_least_BVI}}
\item {\it Gaia} DR2 and EDR3 photometry in the $G$, $G_\mathrm{BP}$ and $G_\mathrm{RP}$ filters \citep{evans2018,riello2021},\footnote{We consider DR2, since DSED provides its 
valuable isochrones only for DR2 but not for EDR3. We compare the colours and magnitudes of the cluster stars, common in DR2 and EDR3, and find no significant systematic difference. 
Accordingly, the DSED isochrones for DR2 are equally suitable for EDR3 and, hence, shown in our CMDs with the EDR3 data. Therefore, we do not present our results obtained with DR2.
Note that the photometry and astrometry of these clusters is exactly the same in {\it Gaia} EDR3 and DR3.}
\item SkyMapper Southern Sky Survey DR3 (SMSS, SMSS DR3) photometry in the $g_\mathrm{SMSS}$, $r_\mathrm{SMSS}$, $i_\mathrm{SMSS}$, and 
$z_\mathrm{SMSS}$ filters \citep{onken2019},\footnote{\url{https://skymapper.anu.edu.au}}
\item $J_\mathrm{VISTA}$ and $Ks_\mathrm{VISTA}$ photometry of the VISTA Hemisphere Survey with the VIRCAM instrument on the Visible and Infrared Survey 
Telescope for Astronomy (VISTA,VHS DR5; \citep{vista}),\footnote{\url{https://cdsarc.cds.unistra.fr/viz-bin/ReadMe/II/367?format=html&tex=true}}
\item {\it Wide-field Infrared Survey Explorer (WISE)} photometry in the $W1$ filter from the unWISE catalogue 
\citep{unwise}.\footnote{\url{https://cdsarc.cds.unistra.fr/viz-bin/cat/II/363}}
\end{enumerate}
Also, we use the following data sets for one of the clusters (see Table~\ref{filters}):
\begin{enumerate}
\item $BV$ photometry of NGC\,6362 with the 2.5 m du Pont telescope of Las Campanas Observatory 
\citep[][hereafter ZKR12]{zloczewski2012},\footnote{\url{https://cdsarc.cds.unistra.fr/viz-bin/cat/J/AcA/62/357}} 
\item $BV$ photometry of NGC\,6362 with the 1 m Swope telescope of Las Campanas Observatory 
\citep[][hereafter NKP17]{narloch2017},\footnote{\url{https://cdsarc.cds.unistra.fr/viz-bin/cat/J/MNRAS/471/1446}}
\item $V$ and Str\"omgren $by$ photometry of NGC\,6723 with the Cerro Tololo Inter-American Observatory (CTIO) 1 m telescope \citep{lee2019}.
\item the fiducial sequences for NGC\,6723 in the $V$, $J_\mathrm{2MASS}$, and $Ks_\mathrm{2MASS}$ filters derived by \citetalias{hendricks2012} from the photometry with the 
New Technology Telescope (NTT) of European Southern Observatory (ESO), La Silla, transformed to the Two Micron All Sky Survey (2MASS) photometric system \citep{2mass}.
\item the fiducial sequences for NGC\,6723 in the $UBVRI$ filters derived by \citet{alcaino1999} from the photometry with the 2.5 m du Pont telescope of Las Campanas Observatory.
Finally, we reject the \citet{alcaino1999}'s data set (see Sect.~\ref{members}) because of its strong systematics due to non-member contamination.
\end{enumerate}

The \citetalias{stetson2019} data sets include photometry from various initial data sets, but not from \citetalias{zloczewski2012}, \citetalias{narloch2017}, or \citet{alcaino1999}.

\begin{table*}
\def\baselinestretch{1}\normalsize\normalsize
\caption[]{The effective wavelength $\lambda_\mathrm{eff}$ (nm), number of stars and median precision of the photometry (mag) for the data sets and filters under consideration.
The number of {\it Gaia} cluster members among the \citetalias{stetson2019} and \citet{lee2019} data sets is shown in brackets.
}
\label{filters}
\[
\begin{tabular}{llccc}
\hline
\noalign{\smallskip}
 Telescope, data set, reference & Filter & $\lambda_\mathrm{eff}$ &  \multicolumn{2}{c}{Number of stars / Median precision} \\
\hline
\noalign{\smallskip}
           &        &                        & NGC\,6362 & NGC\,6723 \\
\hline
\noalign{\smallskip}
{\it HST}/WFC3 \citep{nardiello2018}                              & $F275W$             & 285  & 5692 / 0.02   & 14842 / 0.02    \\
{\it HST}/WFC3 \citep{nardiello2018}                              & $F336W$             & 340  & 7863 / 0.02   & 19886 / 0.02    \\
Various \citepalias{stetson2019}                                  & $U$                 & 366  & 14527 (4651) / 0.02  & 15113 (3309) / 0.02     \\
{\it HST}/WFC3 \citep{nardiello2018}                              & $F438W$             & 438  & 8518 / 0.01   & 20512 / 0.01    \\
{\it HST}/WFPC2 \citep{piotto2002}                                & $F439W$             & 439  & 3275 / 0.04   & 4319 / 0.06     \\
Various \citepalias{stetson2019}                                  & $B$                 & 452  & 18546 (4803) / 0.01  & 22489 (3398) / 0.01    \\
2.5 m du Pont telescope, Las Campanas \citepalias{zloczewski2012} & $B$                 & 452  & 9568 / 0.01   &     --          \\
1 m Swope telescope, Las Campanas \citepalias{narloch2017}        & $B$                 & 452  & 13484 / 0.02  &     --          \\
1 m CTIO telescope \citep{lee2019}                                & Str\"omgren $b$     & 469  &      --       & 27206 (4765) / 0.02    \\
SkyMapper Sky Survey DR3 \citep{onken2019}                        & $g_\mathrm{SMSS}$   & 514  & 2063 / 0.02   & 877 / 0.02       \\ 
{\it Gaia} EDR3 \citep{riello2021}                                & $G_\mathrm{BP}$     & 523  & 5069 / 0.02   & 2207 / 0.03      \\
1 m CTIO telescope \citep{lee2019}                                & Str\"omgren $y$     & 550  &      --       & 27206 (4765) / 0.02    \\
{\it HST}/WFPC2 \citep{piotto2002}                                & $F555W$             & 551  & 3275 / 0.04   & 4319 / 0.05      \\
Various \citepalias{stetson2019}                                  & $V$                 & 552  & 18933 (4803) / 0.01  & 23867 (3398) / 0.01     \\
1 m CTIO telescope \citep{lee2019}                                & $V$                 & 552  &      --       & 27206 (4765) / 0.02     \\
2.5 m du Pont telescope, Las Campanas \citepalias{zloczewski2012} & $V$                 & 552  & 9568 / 0.01   &      --          \\
1 m Swope telescope, Las Campanas \citepalias{narloch2017}        & $V$                 & 552  & 13484 / 0.02  &      --          \\
NTT, ESO, La Silla \citepalias{hendricks2012}                     & $V$                 & 552  &      --       & fiducial / 0.03  \\
{\it HST}/ACS \citep{nardiello2018}                               & $F606W$             & 599  & 21059 / 0.01  & 38689 / 0.01     \\
SkyMapper Sky Survey DR3 \citep{onken2019}                        & $r_\mathrm{SMSS}$   & 615  & 2139 / 0.02   & 977 / 0.02       \\
{\it Gaia} EDR3 \citep{riello2021}                                & $G$                 & 628  & 5069 / 0.01   & 2207 / 0.01      \\
Various \citepalias{stetson2019}                                  & $R$                 & 659  & 16681 (4614) / 0.01  & 21838 (3359) / 0.01     \\
{\it Gaia} EDR3 \citep{riello2021}                                & $G_\mathrm{RP}$     & 770  & 5069 / 0.02   & 2207 / 0.03      \\
SkyMapper Sky Survey DR3 \citep{onken2019}                        & $i_\mathrm{SMSS}$   & 776  & 1545 / 0.02   & 1285 / 0.02      \\
{\it HST}/ACS \citep{nardiello2018}                               & $F814W$             & 807  & 20978 / 0.01  & 38208 / 0.01     \\
Various \citepalias{stetson2019}                                  & $I$                 & 807  & 18698 (4802) / 0.01  & 24356 (3398) / 0.01     \\
SkyMapper Sky Survey DR3 \citep{onken2019}                        & $z_\mathrm{SMSS}$   & 913  & 1202 / 0.02   & 1108 / 0.02      \\
NTT, ESO, La Silla \citepalias{hendricks2012}                     & $J_\mathrm{2MASS}$  & 1234 &    --         & fiducial / 0.03  \\
VISTA VHS DR5 \citep{vista}                                       & $J_\mathrm{VISTA}$  & 1277 & 4401 / 0.05   & 2261 / 0.03      \\
VISTA VHS DR5 \citep{vista}                                       & $Ks_\mathrm{VISTA}$ & 2148 & 2898 / 0.10   & 1927 / 0.09      \\
NTT, ESO, La Silla \citepalias{hendricks2012}                     & $Ks_\mathrm{2MASS}$ & 2176 &      --       & fiducial / 0.03  \\
{\it WISE}, unWISE \citep{unwise}                                 & $W1$                & 3317 &  505 / 0.02   & 504 / 0.01   \\
\hline
\end{tabular}
\]
\end{table*}

Each star has photometric data in some but not all filters. In total, 22 and 26 filters are used for NGC\,6362 and NGC\,6723, respectively. 
They span a wide wavelength range between the UV and middle IR.
For each filter, Table~\ref{filters} indicates the effective wavelength $\lambda_\mathrm{eff}$ in nm, number of stars and the median photometric precision 
after the cleaning of the data set.
Before the cleaning, the initial data sets contained many more stars. The median precision is calculated from the precisions stated by the authors of the data sets.
We use the median precision for calculating predicted uncertainties of the derived [Fe/H], age, distance and reddening, as described in appendix A of \citetalias{ngc6205} 
and Sect.~\ref{results}.

NGC\,6362 and NGC\,6723 are not so rich in photometric data as the GCs from our previous papers. 
However, several twin data sets of accurate photometry are available for both the clusters.
NGC\,6362 and NGC\,6723 are rather similar in their distance from the Sun, age, reddening, [Fe/H], and HB morphology.
Therefore, they are suitable to consider relative estimates of these parameters in the form  `NGC\,6723 minus NGC\,6362', separately derived for each model, 
from an isochrone fitting to each pair of the twin data sets.
Some systematic errors of the models dominate in final absolute estimates of the parameters (see Sect.~\ref{systematics}).
These systematic errors are canceled out in such relative estimates. Hence, we expect the relative estimates to be much more accurate than the absolute ones.

Some initial samples/data sets are selected by use of the VizieR and X-Match services of the Centre de Donn\'ees astronomiques de Strasbourg 
\citep{vizier}\footnote{\url{http://cds.u-strasbg.fr}}.
As described in Sect.~\ref{edr3}, the {\it Gaia} EDR3 data sets are selected within wide areas in order to identify cluster members and derive empirical truncation radii 
of 10 and 8.5 arcmin for NGC\,6362 and NGC\,6723, respectively. These radii are used to select or truncate the other data sets.

Cleaning of the data sets is similar to that done for GCs in \citetalias{ngc288}. 
Typically, we select stars with a photometric error of less than 0.1~mag. However, some data sets have an initial error limit of $<0.1$.
Also the 0.08 mag error limit is applied to the data sets of \citet{nardiello2018} for NGC\,6723 in the $F438W$, $F606W$, and $F814W$ bands, 
{\it Gaia} EDR3, and \citetalias{narloch2017}; the 0.2 mag limit -- to the VISTA $Ks_\mathrm{VISTA}$ photometry.

For the data sets of \citetalias{stetson2019} we select stars with \verb"DAOPHOT" parameters $\chi<3$ and $|{\tt sharp}|<0.3$.
For the {\it HST} WFC3 and ACS photometry, we select stars with $|{\tt sharp}|<0.15$, membership probability $>0.9$ or $-1$, and quality fit $>0.9$.
For the \citetalias{narloch2017} data set, we select stars with a cluster member probability higher than 0.5.
For the \citetalias{zloczewski2012} data set, we select `likely members'. 
For the {\it Gaia} DR2 data sets, we select stars with photometry in all three {\it Gaia} bands and an acceptable parameter 
\verb"phot_bp_rp_excess_factor"$<1.3+0.06$\,\verb"bp_rp"$^2$, as suggested by \citet{evans2018}.

{\it Gaia} EDR3 stars with a precise photometry are selected as those with available data in all three {\it Gaia} bands, with a renormalised unit weight error not exceeding
$1.4$ (\verb"RUWE"$<1.4$), and an acceptable corrected excess factor \verb"phot_bp_rp_excess_factor" (i.e. \verb"E(BP/RP)Corr") between $-0.14$ and $0.14$ \citep{riello2021}.
Unfortunately, this cleaning of the {\it Gaia} EDR3 data sets removes almost all stars in a central arcminute of both the cluster fields. 
The remaining stars at the centres do not show any systematics in CMDs.
Note that the identification of {\it Gaia} cluster members (see Sect.~\ref{edr3}) does not use any photometric data and, hence, is fulfilled before the cleaning of 
the {\it Gaia} EDR3 data sets.

\begin{figure}
\includegraphics{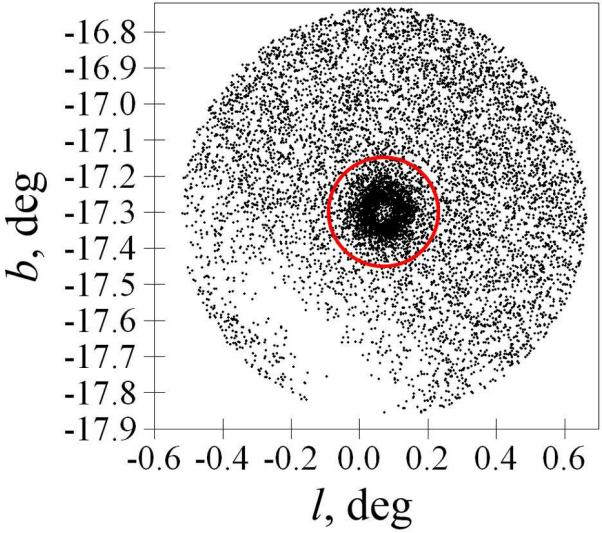}
\caption{The distribution of the initial {\it Gaia} EDR3 data set for NGC\,6723 in the sky in Galactic coordinates within 34 arcmin from the cluster centre.
The red circle shows the selection area within 8.5 arcmin from the cluster centre.
}
\label{ngc6723lb}
\end{figure}

\begin{figure}
\includegraphics{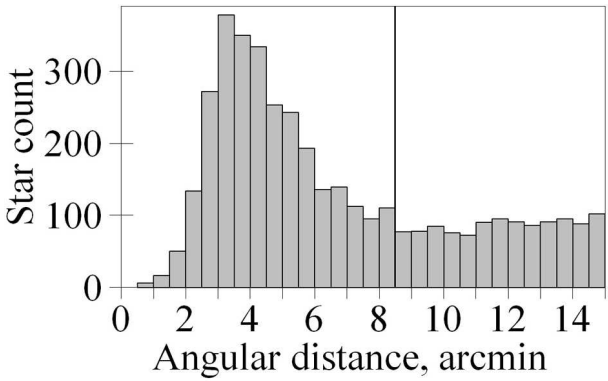}
\caption{The distribution of the initial {\it Gaia} EDR3 data set for NGC\,6723 along the angular distance from the cluster centre.
The vertical line shows the truncation radius of 8.5 arcmin.
}
\label{ngc6723angdist}
\end{figure}

\begin{figure}
\includegraphics{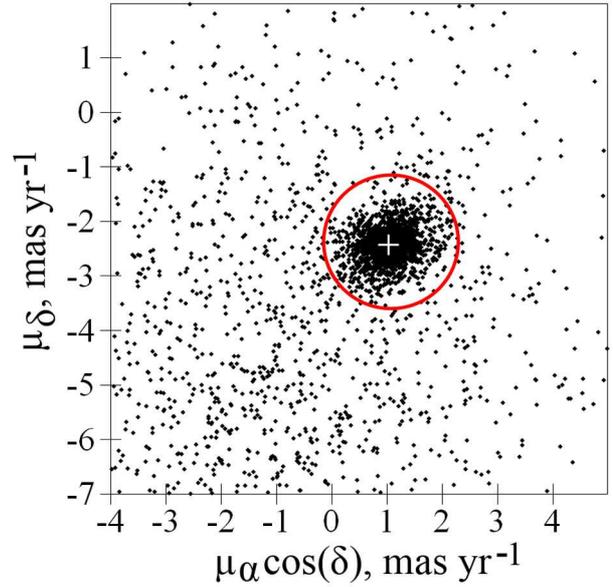}
\caption{The distribution of stars, selected within 8.5 arcmin of the NGC\,6723 centre, over the PM components (mas\,yr$^{-1}$), 
after the remaining cleaning of the sample.
The weighted mean PM and the selection area are shown by the white cross and red circle, respectively.
}
\label{ngc6723mu}
\end{figure}

\begin{figure}
\includegraphics{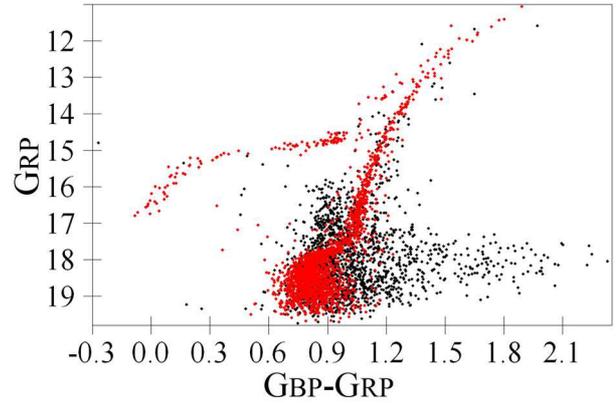}
\caption{$G_\mathrm{BP}-G_\mathrm{RP}$ versus $G_\mathrm{RP}$ CMD for NGC\,6723 stars selected (red symbols) and rejected (black symbols) 
by their parallaxes and PMs after the remaining cleaning of the sample.
}
\label{ngc6723cmd}
\end{figure}

\subsection{{\it Gaia} EDR3 cluster members}
\label{edr3}

Similar to \citetalias{ngc288}, we use accurate PMs and parallaxes from {\it Gaia} EDR3 to select cluster members and calculate systemic cluster PMs and parallaxes. 
Here we briefly summarise this procedure.

First, we select initial {\it Gaia} EDR3 samples within initial radii which are six and four times larger than the halved diameters in Table~\ref{properties} for NGC\,6362 and 
NGC\,6723, respectively. 
The latter is smaller in order to avoid the highly extincted sky region of the Corona Australis cloud complex near NGC\,6723 (see Sect.~\ref{clusters} and \ref{difred}). 
Yet, even within 34 arcmin from the NGC\,6723 centre, the complex strongly obscures the {\it Gaia} EDR3 stars.
This is seen in Fig.~\ref{ngc6723lb} as the void in the distribution of the initial {\it Gaia} EDR3 data set for NGC\,6723.
The red circle shows the finally adopted truncation radius.

Second, the periphery of the initial field is used for estimating the star count surface density of the Galactic background.
Its subtraction allows us to determine the empirical truncation radii of 10 and 8.5 arcmin for NGC\,6362 and NGC\,6723, respectively.
A star count transition from the background to NGC\,6723 at 8.5 arcmin from its centre is presented in Fig.~\ref{ngc6723angdist}.
These truncation radii are very close to the halved diameters presented in Table~\ref{properties}. Few, if any, cluster members should be outside these radii.
Therefore, in order to reduce contamination by non-members, we truncate all the data sets under consideration at these radii (except for data sets with fiducial sequences only).

Third, we reject few stars without PMs and leave only {\it Gaia} EDR3 stars with \verb|duplicated_source|$=0$ (\verb|Dup=0|) and \verb|astrometric_excess_noise|$<1$ ($\epsilon i<1$).
We reject foreground and background stars as those with measured parallax $\varpi>1/R+3\sigma_{\varpi}$ or $\varpi<1/R-3\sigma_{\varpi}$, where $\sigma_{\varpi}$ is the stated 
parallax uncertainty, while $R$ is our estimate of the distance to the cluster.
The latter is upgraded iteratively starting from the \citealt{baumgardt2021} values in Table~\ref{properties}.

Fourth, we adopt initial cluster centre coordinates from \citet{goldsbury2010} and initial systemic PM components 
$\overline{\mu_{\alpha}\cos(\delta)}$ and $\overline{\mu_{\delta}}$ from \citet[][hereafter VB21]{vasiliev2021}.
Then we calculate the standard deviations $\sigma_{\mu_{\alpha}\cos(\delta)}$ and $\sigma_{\mu_{\delta}}$ of the PM components 
$\mu_{\alpha}\cos(\delta)$ and $\mu_{\delta}$ for the cluster members.
We cut off the sample at $3\sigma$, i.e. select cluster members as stars with 
$\sqrt{(\mu_{\alpha}\cos(\delta)-\overline{\mu_{\alpha}\cos(\delta)})^2+(\mu_{\delta}-\overline{\mu_{\delta}})^2}<3\sqrt{\sigma_{\mu_{\alpha}\cos(\delta)}^2+\sigma_{\mu_{\delta}}^2}$.
Then we recalculate the mean coordinates of the cluster centre and weighted mean systemic PM components.
This procedure is repeated iteratively until we stop losing stars in the $3\sigma$ cut.

Note that since the stated {\it Gaia} EDR3 PM uncertainty increases strongly with magnitude, faint cluster members make a negligible contribution
to the weighted mean systemic PMs.

The final empirical standard deviations $\sigma_{\mu_{\alpha}\cos(\delta)}$ and $\sigma_{\mu_{\delta}}$ are reasonable, being slightly higher than the mean stated PM uncertainties: 
0.25 versus 0.17 and 0.34 versus 0.27 mas\,yr$^{-1}$ for NGC\,6362 and NGC\,6723, respectively (averaged for the PM components).
Fig.~\ref{ngc6723mu} shows a distribution of stars, selected within 8.5 arcmin of the NGC\,6723 centre, over the PM components, after the remaining cleaning of the sample 
(see Sect.~\ref{datasets}). 
The cluster members are those inside the red circle. The CMD of the stars from Fig.~\ref{ngc6723mu} is shown in Fig.~\ref{ngc6723cmd}.

\begin{table}
\def\baselinestretch{1}\normalsize\normalsize
\caption[]{The cluster systemic PMs (mas\,yr$^{-1}$). The random uncertainties are presented for the PMs from this study and from \citet{vitral2021},
while the total (random plus systematic) uncertainty is presented for the PMs from \citetalias{vasiliev2021}.
}
\label{systemic}
\[
\begin{tabular}{llcc}
\hline
\noalign{\smallskip}
Cluster & Source & $\mu_{\alpha}\cos(\delta)$ & $\mu_{\delta}$ \\
\hline
\noalign{\smallskip}
          & This study                & $-5.512\pm0.003$ & $-4.780\pm0.004$ \\
NGC\,6362 & \citetalias{vasiliev2021} & $-5.504\pm0.024$ & $-4.763\pm0.024$ \\
          & \citet{vitral2021}        & $-5.509\pm0.003$ & $-4.763\pm0.003$ \\
\noalign{\smallskip}
          & This study                & $1.021\pm0.008$ & $-2.427\pm0.007$ \\
NGC\,6723 & \citetalias{vasiliev2021} & $1.030\pm0.026$ & $-2.418\pm0.026$ \\
          & \citet{vitral2021}        & $1.028\pm0.006$ & $-2.419\pm0.006$ \\
\hline
\end{tabular}
\]
\end{table}

Table~\ref{systemic} presents our final weighted mean PMs in comparison to those from \citetalias{vasiliev2021} and \citet{vitral2021}.
They are also derived from {\it Gaia} EDR3, but using different approaches.
The estimates are consistent within $\pm0.01$ mas\,yr$^{-1}$.
However, all these estimates may be equally affected by yet poorly known {\it Gaia} EDR3 PM systematic errors.
\citetalias{vasiliev2021} note that such errors impose `the irreducible systematic floor' on the total accuracy of the {\it Gaia} EDR3 PMs 
`for any compact stellar system'. 
Such total uncertainty is given in Table~\ref{systemic} for the estimates of \citetalias{vasiliev2021},
while only the random uncertainty is given for our and \citet{vitral2021} estimates. We adopt the total uncertainties as the final ones of our PMs.

Similarly, the total uncertainty of {\it Gaia} EDR3 parallaxes, determined by \citetalias{vasiliev2021} as 0.011 mas, is adopted for our median 
parallaxes of cluster members.
After the correction of the parallax zero-point following \citet{lindegren2021b}, we present these median parallaxes in Table~\ref{properties}.
We compare these parallaxes with distances, derived in our isochrone fitting, in Sect.~\ref{distance}.

The final lists of the {\it Gaia} EDR3 cluster members are presented in Table~\ref{gaiaedr3}.

\begin{table}
\def\baselinestretch{1}\normalsize\normalsize
\caption[]{The list of the {\it Gaia} EDR3 members of NGC\,6362 and NGC\,6723. The complete table is available online.
}
\label{gaiaedr3}
\[
\begin{tabular}{cc}
\hline
\noalign{\smallskip}
NGC\,6362 & NGC\,6723 \\
\hline
\noalign{\smallskip}
5813076160252430208 & 6730890124282434688 \\
5813077259764184320 & 6730890193014007040 \\
5813077264076635648 & 6730890880196684672 \\
5813077465922524416 & 6730890880208577280 \\
5813077500282241024 & 6730890948916284544 \\
\ldots & \ldots \\
\hline
\end{tabular}
\]
\end{table}

\begin{figure}
\includegraphics{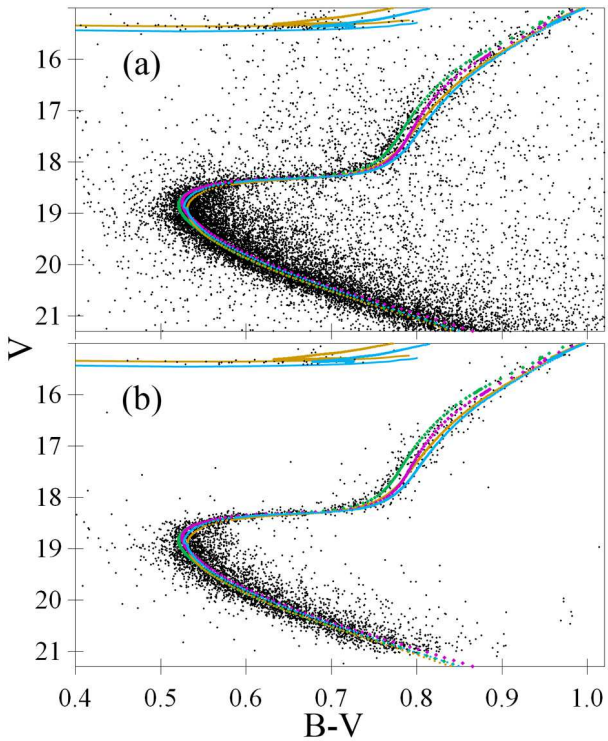}
\caption{$B-V$ versus $V$ CMDs of NGC\,6362 for (a) the initial \citetalias{stetson2019} data set and (b) the {\it Gaia} EDR3 cluster members from the \citetalias{stetson2019} data set.
The isochrones (same for both the charts) from BaSTI for $Y=0.25$ (blue) and 0.275 (brown) and from DSED for $Y=0.25$ (purple) and 0.275 (green) are calculated with the best-fitting 
parameters from Table~\ref{cmds}.
}
\label{ngc6362_stetson_full_gaia}
\end{figure}

\subsection{Cluster members in other data sets}
\label{members}

NGC\,6362 and NGC\,6723 are at middle Galactic latitudes and their fields are moderately contaminated by foreground stars.
Therefore, special effort is needed to identify cluster members (except data sets with fiducial sequences only).
\citet{nardiello2018}, \citet{piotto2002}, \citetalias{zloczewski2012}, and \citetalias{narloch2017} have cleaned their data sets from non-members by use of dedicated PMs. 
Although imperfect, this membership identification cannot be significantly improved by use of the {\it Gaia} data.
In contrast, we create SMSS, VISTA, and unWISE data sets for our study by cross-identification of our {\it Gaia} EDR3 cluster members with the SMSS,
VISTA, and unWISE catalogues, respectively.

The third group, the data sets of \citetalias{stetson2019} and \citet{lee2019}, can be fitted by isochrones as is. 
However, we cross-identify our {\it Gaia} EDR3 cluster members with these data sets and fit isochrones to both the initial data sets and to {\it Gaia} EDR3 cluster members in 
these data sets.
The {\it Gaia} membership identification appears to be important, since it not only cleans fiducial sequences but also correct them for a bias due to a non-uniform distribution 
of non-members in CMDs.
Note that Table~\ref{filters} shows a majority of the \citetalias{stetson2019} and \citet{lee2019} initial stars lost in this cross-identification with {\it Gaia}. 
However, many lost stars are not non-members but stars, which are too faint for {\it Gaia}.
Accordingly, they are rather faint MS stars, that cannot impact our results.

Fig.~\ref{ngc6362_stetson_full_gaia} shows an initial \citetalias{stetson2019} CMD versus the same CMD with the {\it Gaia} EDR3 cluster members from the \citetalias{stetson2019} 
data set.
It is seen that the {\it Gaia} EDR3 cluster members are limited to about 2 mag fainter than TO.
The initial data set has a lot of contaminants around the HB, above the SGB and redder than the MS.
They would bias the best-fitting isochrones and, hence, the derived parameters of the cluster.
Use of the initial data set instead of its {\it Gaia} EDR3 cluster members would increase derived distance by about 50 pc. 
Age derived from the HB-SGB magnitude difference would decrease by about 0.5 Gyr, while age derived from the SGB length would increase by about 0.5 Gyr.
It is seen that the isochrones in Fig.~\ref{ngc6362_stetson_full_gaia}~(a) tend to be redder than the contaminated RGB and bluer than the contaminated MS, 
while they fit the clean RGB and MS in Fig.~\ref{ngc6362_stetson_full_gaia}~(b) much better.
As a result, contaminants would decrease/increase the reddening derived for the RGB/MS, respectively, by few hundredths of a magnitude.
Thus, contaminants introduce a discrepancy between the cluster parameters derived from different domains of CMD.
Hence, the final results from contaminated data would depend on our preferences to use one domain or another.
Finally, hereafter, we use only the {\it Gaia} EDR3 cluster members from the \citetalias{stetson2019} and \citet{lee2019} data sets instead of the initial data sets.
The expense is that we lose the faint MS stars and almost all stars within a central arcminute of the clusters.

An example of a bias of a derived parameter due to a contamination of a CMD is the \citet{alcaino1999} data set.
We do not have its star-by-star photometry, but only its fiducial sequences.
Hence, we cannot cross-identify it and eliminate non-members. A strong non-member contamiation of the MS is seen in \citet{alcaino1999}'s figures.
As a result, the median $B-V$ TO colour of the \citet{alcaino1999} data set is about 0.04 mag redder than that of \citetalias{stetson2019} data set.
Therefore, we decided to reject the \citet{alcaino1999} data set.

Fig.~\ref{ngc6362_stetson_full_gaia} shows a segregation of the RGB into two populations. The redder population is better fitted by the BaSTI isochrone with $Y=0.25$
(though the DSED isochrone with $Y=0.25$ is also acceptable), while the bluer population -- by the DSED isochrone with $Y=0.275$. 
This suggests a helium abundance difference $\Delta Y\approx0.025$ between the populations, in line with its estimates in Sect.~\ref{clusters}.
Note that the colour difference between the BaSTI isochrones with $Y=0.25$ and 0.275 is much lower than that between the observed segregated RGBs.
However, the observed segregated AGBs are well reproduced by these BaSTI isochrones.

\begin{table}
\def\baselinestretch{1}\normalsize\normalsize
\caption[]{The fiducial sequences for NGC\,6362 and NGC\,6723 $G_\mathrm{RP}$ versus $G_\mathrm{BP}-G_\mathrm{RP}$ based on the data of
{\it Gaia} EDR3. The complete table is available online.
}
\label{fiducial}
\[
\begin{tabular}{cccc}
\hline
\noalign{\smallskip}
\multicolumn{2}{c}{NGC\,6362} & \multicolumn{2}{c}{NGC\,6723} \\
\noalign{\smallskip}
$RP$ & $BP-RP$ & $RP$ & $BP-RP$ \\
\hline
\noalign{\smallskip}
15.48 & 0.12 & 16.75 & -0.07 \\
15.40 & 0.14 & 16.50 & -0.02 \\
15.34 & 0.16 & 15.94 & 0.07 \\
15.24 & 0.20 & 15.67 & 0.13 \\
15.20 & 0.22 & 15.40 & 0.20 \\
\ldots & \ldots & \ldots & \ldots \\
\hline
\end{tabular}
\]
\end{table}

\begin{figure}
\includegraphics{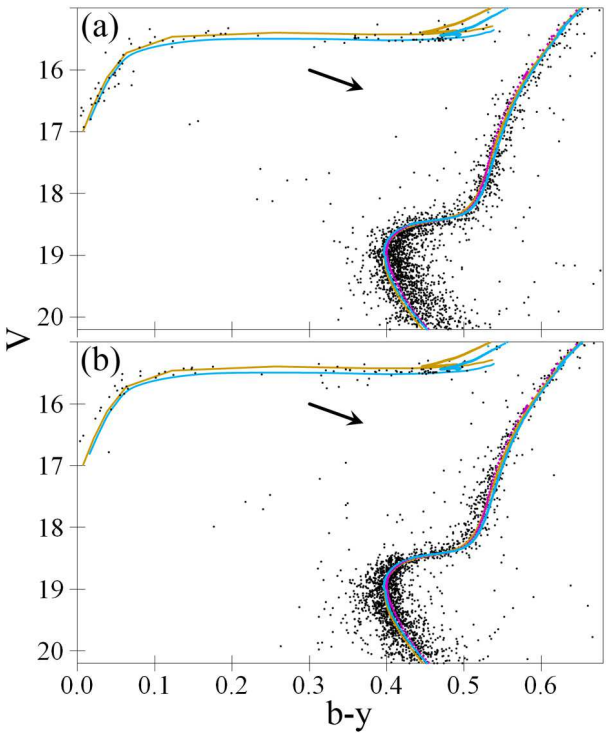}
\caption{A part of the $b-y$ versus $V$ CMD for the data from \citet{lee2019} for the (a) southern and (b) northern half (in Galactic coordinates) of the NGC\,6723 field.
The isochrones (same for both the charts) from BaSTI for $Y\approx0.25$ (blue) and 0.275 (brown) and from DSED for $Y\approx0.26$ (purple) are calculated with the best-fitting 
parameters from Table~\ref{cmds}. The black arrow shows reddening and extinction corresponding to $E(B-V)=0.1$ mag.
}
\label{ngc6723difred_lee2019}
\end{figure}

\begin{figure}
\includegraphics{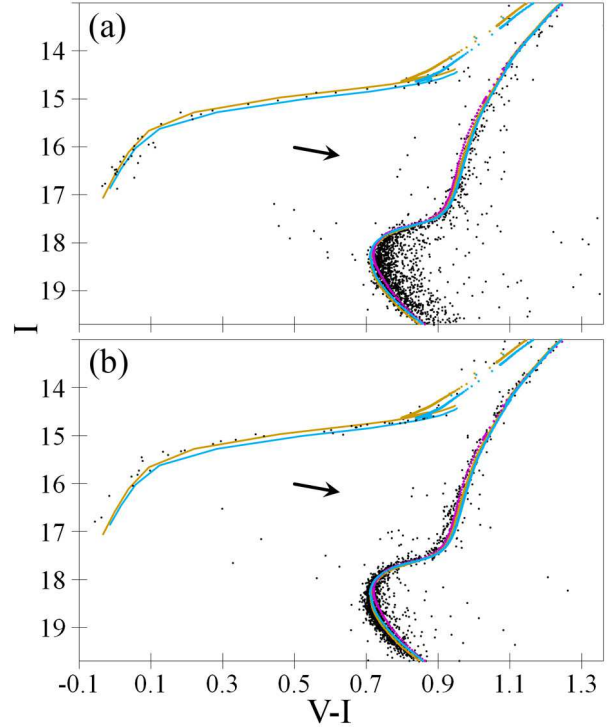}
\caption{The same as Fig.~\ref{ngc6723difred_lee2019} but for $V-I$ versus $I$ CMD for the {\it Gaia} EDR3 cluster members among the \citetalias{stetson2019} data set for NGC\,6723.
}
\label{ngc6723difred_stetson2019}
\end{figure}

\begin{figure}
\includegraphics{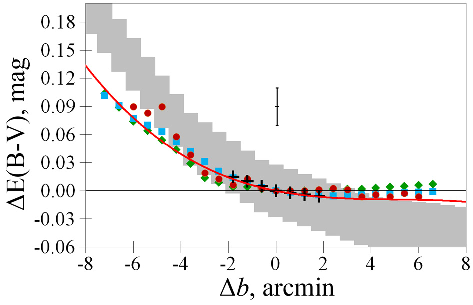}
\caption{Differential reddening $\Delta E(B-V)$ in the field of NGC\,6723 as a function of Galactic latitude $\Delta b$ w.r.t. the cluster centre: derived from the 
$b-y$ versus $V$ CMD \citep{lee2019} -- brown circles, 
$B-V$ versus $V$ CMD \citepalias{stetson2019} -- green diamonds,
$V-I$ versus $I$ CMD \citepalias{stetson2019} -- blue squares, and
{\it HST} $F606W-F814W$ versus F814W CMD \citep{nardiello2018} -- black crosses
by use of \citetalias{ccm89} extinction law with $R_\mathrm{V}=3.1$.
The uncertainty of $\pm0.02$ mag for these results is shown by the separate error bar. 
The differential reddening from \citetalias{sfd98} with its stated precision $\pm0.028$ mag is shown by the grey area.
The approximating equation~\ref{difredequ} is shown by the red curve.
}
\label{ngc6723difred}
\end{figure}

\begin{figure}
\includegraphics{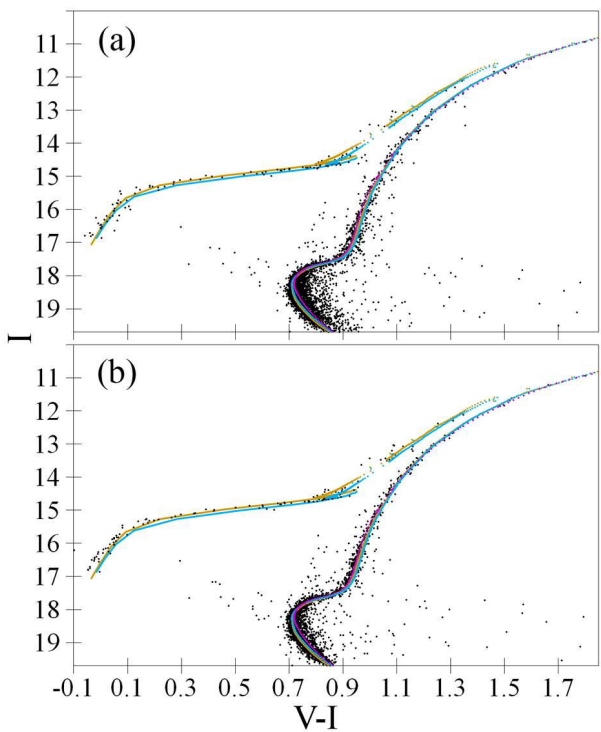}
\caption{The same as Fig.~\ref{ngc6723difred_stetson2019} but for the whole NGC\,6723 field (a) before and (b) after the correction for differential reddening.
}
\label{ngc6723_stetson_vi}
\end{figure}

\subsection{Differential reddening}
\label{difred}

The differential reddening in the field of NGC\,6723 forces us to correct it for all data sets, except \citetalias{hendricks2012} and \citet{piotto2002}. 
Fortunately, the differential reddening in the NGC\,6723 field increases along Galactic latitude, since its apparent source, the Corona Australis cloud complex, 
is located south of NGC\,6723 (in Galactic coordinates). \citetalias{sfd98} and our CMDs confirm a differential reddening gradient with latitude.
Two examples of such CMDs are shown in Figs~\ref{ngc6723difred_lee2019} and \ref{ngc6723difred_stetson2019}.
Stars of the southern and northern (in Galactic coordinates) halves of the NGC\,6723 field tend to be on the opposite sides of the same reference isochrones, 
though the southern stars show a larger deviation and scatter. 
The reddening vector indicates that the southern stars are shifted w.r.t. the northern ones by a higher reddening of the order of $E(B-V)\approx0.1$ mag.
This effect is the same for all CMD domains.

We calculate differential reddening as a function of latitude using the approach applied in \citetalias{ngc6205} and \citetalias{ngc288}.
Namely, some partial fiducial sequences for several latitude bins are analyzed.
Only the large data sets of \citet{nardiello2018}, {\it Gaia} EDR3, \citetalias{stetson2019}, and \cite{lee2019} have enough stars to consider differential reddening in many 
latitude bins.
For comparison, the obtained differential reddenings are set to zero at the cluster centre and converted into $\Delta E(B-V)$ by use of the \citet[][hereafter CCM89]{ccm89} 
extinction law with extinction-to-reddening ratio $R_\mathrm{V}\equiv A_\mathrm{V}/E(B-V)=3.1$.
Fig.~\ref{ngc6723difred} shows the differential reddening in the field of NGC\,6723 as a function of Galactic latitude for the data sets of \citet{lee2019} ($b-y$ versus $V$ CMD), 
\citetalias{stetson2019} ($B-V$ versus $V$ and $V-I$ versus $I$ CMDs), and \citet{nardiello2018} ($F606W-F814W$ versus $F814W$ CMD).
Also, the differential reddening from \citetalias{sfd98} is shown by the grey area. Good agreement between all data sets is observed.

Note that the {\it HST} data set of \citet{nardiello2018} covers only a central area of $3.4\times3.4$ arcmin and, accordingly, shows a differential reddening corresponding to 
only $\Delta E(B-V)=0.02$ mag over the area.
Such negligible differential reddening must appear for any data set covering only a central part of the field\footnote{To our knowledge, this applies to the 
\citetalias{hendricks2012} and \citet{piotto2002} data sets.}. This may explain why such a strong differential reddening at the periphery was not recognized earlier.

A deviation of the data sets from the \citetalias{sfd98} in Fig.~\ref{ngc6723difred} can be explained by (i) systematic variations of the cluster stellar content with its radius or 
(ii) \citetalias{sfd98} calibration errors. Either way, we assume a decrease of reddening to the North across the whole cluster field. 
Therefore, we approximate the differential reddening by a third-order polynomial taking into account the differential reddening results from all the large data sets:
\begin{equation}
\label{difredequ}
\Delta E(B-V)=-13.6750\Delta b^3+3.5958\Delta b^2-0.30\Delta b,
\end{equation}
where $\Delta b$ is the latitude offset in degrees w.r.t. the cluster centre.
This polynomial is shown in Fig.~\ref{ngc6723difred} by the red curve. This suggests a differential reddening up to $\Delta E(B-V)\approx0.13$ mag at
the Southern periphery of the NGC\,6723 field.

We use equation~(\ref{difredequ}) to correct the differential reddening in all the data sets, except \citetalias{hendricks2012} and \citet{piotto2002}.
This procedure reduces the scatter of the photometric data around the fiducial sequences.
An example is given in Fig.~\ref{ngc6723_stetson_vi} where the CMD from Fig.~\ref{ngc6723difred_stetson2019} is shown for the whole NGC\,6723 field
(a) before and (b) after the correction for differential reddening.

All reddening estimates for NGC\,6723, obtained in our isochrone fitting of CMDs, are referred to the cluster centre, where reddening and extinction are rather low.
Hence, these are not estimates of an average or median reddening in the NGC\,6723 field.
Therefore, to estimate reddening in a part of the NGC\,6723 field, one should estimate differential reddening in this part by use of equation~(\ref{difredequ}) and add it 
to our reddening estimate for the cluster centre.

\begin{figure}
\includegraphics{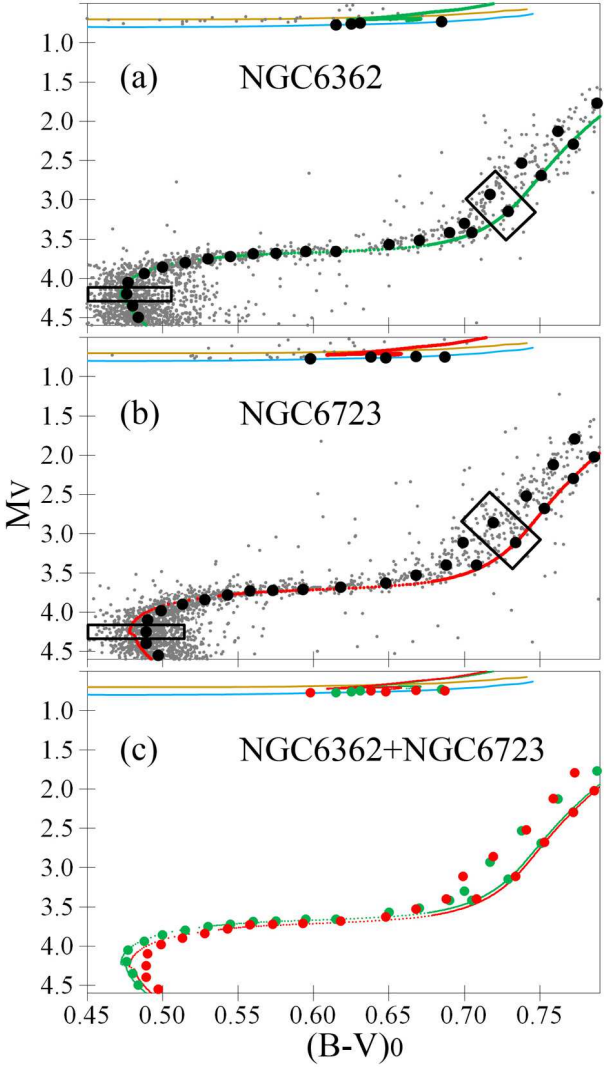}
\caption{Hertzsprung--Russell diagram $(B-V)_0$ versus $M_\mathrm{V}$ for NGC\,6362 and NGC\,6723 with the {\it Gaia} EDR3 members of the \citetalias{stetson2019} data sets:
stars - grey dots, fiducial points - black circles (green and red in the bottom plot for NGC\,6362 and NGC\,6723, respectively), 
best fitting BaSTI ZAHB for [Fe/H]$=-1.05$ and $Y=0.25$ - blue curve, best fitting BaSTI ZAHB for [Fe/H]$=-1.05$ and $Y=0.275$ - brown curve,
NGC\,6362 best fitting BaSTI isochrone for [Fe/H]$=-1.05$, $Y=0.25$ and age 12 Gyr - green curve,
NGC\,6723 best fitting BaSTI isochrone for [Fe/H]$=-1.05$, $Y=0.25$ and age 12.5 Gyr - red curve.
All the data are corrected for reddening, extinction, and distance estimates from Table~\ref{cmds}. 
Black boxes show the cross sections from Fig.~\ref{cross}.
}
\label{cmd1}
\end{figure}

\begin{figure}
\includegraphics{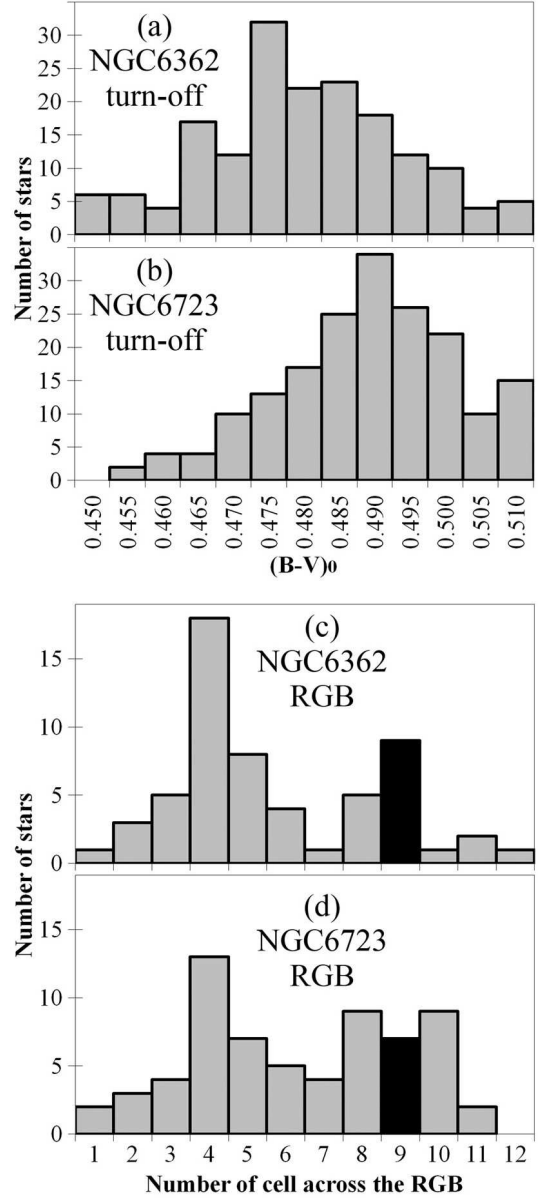}
\caption{The distribution of the TO stars along $(B-V)_0$ and RGB stars across the RGB inside the boxes from Fig.~\ref{cmd1}.
The black bars mark the positions of the RGB fiducial sequence points.
}
\label{cross}
\end{figure}

\subsection{Fiducial sequences}
\label{fiducials}

In order to fit data by a theoretical isochrone, we present them by a fiducial sequence, i.e. a colour--magnitude relation for single stars.
We calculate such a sequence as a locus of the number density maxima in some colour--magnitude bins. Some details and examples are given in 
section 3 of \citetalias{ngc6205}.

The HB, AGB, bright RGB and sometimes other CMD domains contain a small number of stars.
In such a case, a fiducial point is defined by a few or even just one star with rather precise photometry, i.e. if a colour and magnitude of such fiducial point/points can 
be defined within $\pm0.04$ mag.
The balance of uncertainties, presented in appendix A of \citetalias{ngc6205}, shows that such an uncertainty of a fiducial point is negligible w.r.t.
total uncertainty, since we use a lot of fiducial points to derive [Fe/H], distance, age, and reddening.

The fiducial sequences for the {\it Gaia} EDR3 data sets are presented in Table~\ref{fiducial} as an example.
All other fiducial sequences can be provided on request.

We find that the populations are segregated at the HB in almost all CMDs, at the AGB in CMDs with a considerable number of the AGB stars,
as well as at the RGB in some CMDs (an example is gives in Fig.~\ref{ngc6362_stetson_full_gaia}).
We find no CMD for these GCs with the populations segregated at the SGB, TO, or MS.

In the case of no segregation, an unresolved mix of the populations is presented by a fiducial sequence, which, in turn, is fitted by an interpolated 
isochrone with $Y=0.26$, as noted in Sect.~\ref{iso}.
In the case of segregation, each population is presented by its own fiducial sequence, which, in turn, is fitted by its own primordial or helium-enriched 
isochrone of the same model, [Fe/H], distance, reddening, and age.
However, in some cases, we cannot fit a population by an isochrone with a reasonable abundance.

To derive the best-fitting parameters from a CMD, we select an isochrone with a minimal total offset between the isochrone points 
and the fiducial points in the same magnitude range of the CMD.

Fig.~\ref{cmd1} shows an example of isochrones fitting fiducial sequences, which, in turn, are constructed for bulks of stars.
This figure combines two CMDs for the twin \citetalias{stetson2019} data sets.
All 9 CMDs of the twin data sets provide a similar gauge, since they show similar results (see Table~\ref{cmds}).
When comparing two clusters, we present this as the Hertzsprung--Russell diagrams created from related CMDs 
(one of which is presented in Fig.~\ref{ngc6362_stetson_full_gaia}) by use of reddening, extinction, 
and distance estimates from Table~\ref{cmds} and the same [Fe/H]$=-1.05$.
It is worth noting that a good fitting of the fiducial points by the isochrones in Fig.~\ref{cmd1}, when the HB and RGB of two clusters coincide, 
indicates that we use correct reddenings, extinctions, and distances for both the clusters.

Example areas across the TO and RGB are shown by the black boxes in Fig.~\ref{cmd1}. Their profiles are given in Fig.~\ref{cross}.
A median colour can be easily derived from the TO profiles. 
In contrast, the RGB profiles show the segregation of stars, discussed in Sect.~\ref{members} and seen in Fig.~\ref{ngc6362_stetson_full_gaia}.
The bluer population (presumably helium-enriched) dominates. However, the stars counts allow us to set fiducial sequence points with a precision
of a hundredth of a magnitude for colour and about 0.1 -- for magnitude, as seen in Fig.~\ref{cmd1}.

Fig.~\ref{cmd1} shows that each HB fiducial point is defined by only one star. Yet, the {\it Gaia} membership cross-identification 
(see Sect.~\ref{members}) ensures us that such a star is a cluster member.
HB stars magnitude better defines distance modulus of clusters. Rather bright HB stars in the \citetalias{stetson2019} data sets have
a magnitude uncertainty of about 0.01 mag. Therefore, even a few engaged HB stars provide us with a statistical uncertainty of an average or median HB 
magnitude of better than 0.01 mag, i.e. much better than a systematic uncertainty of about 0.04 mag (see Sect.~\ref{systematics}).
Note that the HB stars, which are bluer than those marked by the fiducial points in Fig.~\ref{cmd1}, can be RR~Lyrae variables or can belong to a helium-enriched 
population and, hence, must be fitted by a helium-enriched isochrone (for example, by that with $Y=0.275$ in Fig.~\ref{cmd1}).

Similar to the HB, the SGB magnitude and the HB--SGB magnitude difference can be determined with a random precision of about 0.01 mag and
systematic accuracy of few hundredths of a magnitude. 
The HB--SGB magnitude difference better defines cluster age.
The high precision of the {\it Gaia}--\citetalias{stetson2019} data sets allows one to see in Fig.~\ref{cmd1}~(c) that the SGB of NGC\,6362 is 
about 0.04 mag brighter than that of NGC\,6723. This means that the former is slightly younger than the latter: 
the best-fitting isochrones in Fig.~\ref{cmd1}~(c) differ by 0.5 Gyr.
It would seem that the magnitude difference of 0.04 mag can easily get lost in the systematic uncertainties of 0.04 for both the HB and SGB magnitudes.
Moreover, the situation in Fig.~\ref{cmd1}~(c) is further aggravated by the segregation of the faint RGB.
However, the systematic uncertainties must be the same for both clusters and, hence, are canceled in relative estimates.
Yet, an age difference of about 0.5 Gyr may be the sensitivity limit for such an isochrone fitting with the best current data sets and models.

Age is also defined by the length of the SGB between the TO and RGB. 
Fig.~\ref{cmd1}~(c) shows that NGC\,6362 has a longer SGB and, hence, is younger. Thus, two proxies of age agree in this case.
However, the length of the SGB may be affected by significant systematics from stellar modeling.
This appears to be a bad fit of the TO when the HB and RGB of two clusters coincide.
This is seen for NGC\,6723 in Fig.~\ref{cmd1}~(c).

\begin{table*}
\def\baselinestretch{1}\normalsize\normalsize
\caption{The results of the isochrone fitting for various models and some key CMDs for both the clusters. In all the CMDs, the colour is the abscissa and the magnitude in the 
redder filter is the ordinate, except the \citet{lee2019} CMD where $b-y$ is the abscissa and $V$ is the ordinate. The NGC\,6723 reddenings are reffered to the cluster centre.
The derived reddenings are converted to $E(B-V)$, given in parentheses, by use of extinction coefficients from \citet{casagrande2014,casagrande2018a,casagrande2018b}.
The complete table is available online.
}
\label{cmds}
\[
\begin{tabular}{lcccc}
\hline
\noalign{\smallskip}
 & \multicolumn{2}{c}{NGC\,6362} & \multicolumn{2}{c}{NGC\,6723} \\
\noalign{\smallskip}
CMD colour                                  & DSED &  BaSTI & DSED &  BaSTI  \\
\hline
$E(F606W-F814W)$ {\it HST} ACS              & $0.074\pm0.02$ (0.08) & $0.058\pm0.02$ (0.06) & $0.079\pm0.03$ (0.08) & $0.060\pm0.03$ (0.06) \\
$F606W-F814W$ age, Gyr	                    & 12.5            & 12.5           & 12.5           & 13.0           \\
$F606W-F814W$ distance, kpc                 & 7.7             & 7.7            & 8.2            & 8.1            \\
$F606W-F814W$ [Fe/H]	                    & $-1.10$         & $-1.00$        & $-1.15$        & $-1.00$        \\
\noalign{\smallskip}
$E(G_\mathrm{BP}-G_\mathrm{RP})$ EDR3       & $0.132\pm0.03$ (0.10) & $0.093\pm0.03$ (0.07) & $0.126\pm0.03$ (0.10) & $0.097\pm0.03$ (0.07) \\
$G_\mathrm{BP}-G_\mathrm{RP}$ age, Gyr      &   12.5         & 12.5           & 13.0           & 13.0           \\
$G_\mathrm{BP}-G_\mathrm{RP}$ distance, kpc &   7.6          & 7.5            & 8.1            & 8.1            \\
$G_\mathrm{BP}-G_\mathrm{RP}$ [Fe/H]        & $-1.15$        & $-1.00$        & $-1.15$        & $-1.05$        \\
\noalign{\smallskip}
$E(B-V)$ \citetalias{stetson2019}           & $0.071\pm0.03$ (0.07) & $0.053\pm0.03$ (0.05) & $0.085\pm0.03$ (0.09) & $0.060\pm0.03$ (0.06) \\
$B-V$ age, Gyr                              & 11.5           & 12.0           & 12.5           & 12.5           \\
$B-V$ distance, kpc                         & 7.8            & 7.8            & 7.9            & 8.1            \\
$B-V$ [Fe/H]                                & $-1.10$        & $-1.05$        & $-1.15$        & $-1.05$        \\
\noalign{\smallskip}
$E(g-r)$ SkyMapper DR3                      & $0.047\pm0.02$ (0.07) & $0.040\pm0.02$ (0.06) & $0.059\pm0.03$ (0.08) & $0.048\pm0.03$ (0.07) \\
$g-r$ age, Gyr                              & 12.0            & 12.0           & 12.0           & 12.5           \\
$g-r$ distance, kpc                         & 7.7             & 7.7            & 8.3            & 8.2            \\
$g-r$ [Fe/H]                                & $-1.10$         & $-1.00$        & $-1.15$        & $-1.05$ \\
\noalign{\smallskip}
$E(F439W-F555W)$ {\it HST} WFPC2            & $0.056\pm0.03$ (0.05) & $0.041\pm0.03$ (0.04) & $0.067\pm0.03$ (0.07) & $0.046\pm0.03$ (0.04) \\
$F439W-F555W$ age, Gyr	                    & 12.5            & 12.0           & 12.0           & 12.5           \\
$F439W-F555W$ distance, kpc                 & 7.9             & 8.1            & 8.6            & 8.5            \\
$F439W-F555W$ [Fe/H]	                    & $-1.10$         & $-1.05$        & $-1.15$        & $-1.05$        \\
\ldots & \ldots & \ldots & \ldots & \ldots \\
\hline
\end{tabular}
\]
\end{table*}

\begin{figure}
\includegraphics{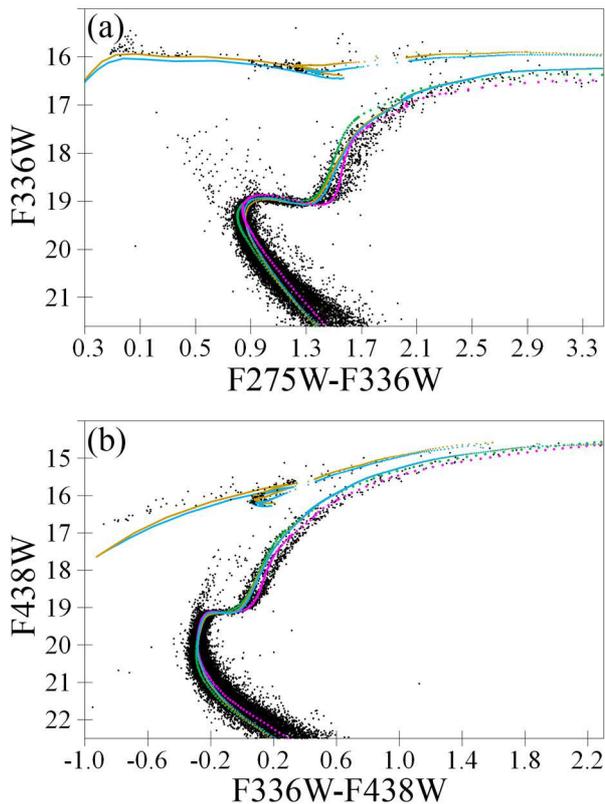}
\caption{The UV CMDs for NGC\,6723 with the (a) $F275W-F336W$ and (b) $F336W-F438W$ colours. The isochrones from BaSTI for $Y\approx0.25$ (blue) and 0.275 (brown) and from 
DSED for $Y\approx0.25$ (purple) and 0.33 (green) are calculated with the best-fitting parameters from Table~\ref{cmds}.
}
\label{ngc6723_wfc3}
\end{figure}

\section{Results}
\label{results}

With this wealth of photometric data, we fit isochrones to dozens of CMDs with different colours.
As in our previous papers, results for adjacent CMDs are consistent within their precision: e.g. results for a CMD with the $B-I$ colour are consistent with those for CMDs with 
the $B-V$, $V-R$, and $R-I$ colours.
Hence, we show only some examples of the CMDs with isochrone fits in Figs~\ref{ngc6723_stetson_vi} and \ref{ngc6723_wfc3}.
Some more examples are given in Appendix~\ref{addcmds}.
Figures for all the CMDs can be provided on request. The derived [Fe/H], ages, distances, and reddenings for the most important CMDs are presented in Table~\ref{cmds}.
For comparison, the derived reddenings are converted to $E(B-V)$, given in parentheses, by use of extinction coefficients from \citet{casagrande2014,casagrande2018a,casagrande2018b}.

For most CMDs, the isochrone-to-fiducial fitting is so precise that the best-fitting isochrones of the models almost coincide with each other and with our fiducial 
sequence on the scales of our CMD figures for the RGB, SGB, and TO. Hence, usually we do not show our fiducial sequences for clarity.

The predicted statistic uncertainties of the fitting are similar to those described in the balance of uncertainties, presented in appendix A of \citetalias{ngc6205}.
For each combination of a fiducial sequence and its best-fitting isochrone we find the maximal offset of this isochrone w.r.t. 
this fiducial sequence along the reddening vector (i.e. nearly along the colour). 
Such an offset is given in Table~\ref{cmds} after each value of reddening as its empirical uncertainty.
Usually, the predicted and empirical uncertainties are comparable. The largest value in such a pair of the uncertainties is shown by 
error bars in Figs~\ref{ngc6362law} and \ref{ngc6723law}, which demonstrate our resulting extinction laws.
However, it is worth noting that systematic uncertainties of isochrones always dominate in final uncertainty (see Sect.~\ref{systematics}).

For example, let us consider the uncertainty of [Fe/H] derived from the slope of the RGB.
Taking into account RGB geometry, the number of stars used, their photometric precision, and slope sensitivity to variations of [Fe/H], we predict a random uncertainty of [Fe/H], 
derived from a pair of a CMD and a model, as of about 0.15 dex. 
Eleven (NGC\,6362) or ten (NGC\,6723) CMDs, used for our final average [Fe/H] estimate, might provide their uncertainty of about 0.05 dex 
by use of a model. However, the models may show a systematic difference between their average [Fe/H] estimates larger than 0.05 dex.

\subsection{Issues}
\label{issues}

The VISTA photometry strongly deviates blueward from the isochrones for the bright RGB ($J_\mathrm{VISTA}<12$ and $<13$ for NGC\,6362 and NGC\,6723, respectively). 
It seems to be an error in the VISTA photometry for very bright stars. However, the remaining VISTA stars are enough to derive accurate parameters.

Generally, an UV, UV--optical, optical--IR, and IR--IR CMD contains less stars, more contaminants, and less populated HB, TO, and MS than a typical optical CMD. 
Moreover, isochrones are less accurate for such CMDs. 
In addition, {\it HST} WFC3 UV filters are sensitive to the CNO abundances through OH, CN, NH, and CH bands \citep{vandenberg2022}.
However, some of such CMDs provide reliable [Fe/H], distance, age, and reddening estimates (see Table~\ref{cmds}).
Examples, given in Fig.~\ref{ngc6723_wfc3}, show that almost all UV CMD domains are fitted by the isochrones rather successfully, i.e. with
reasonable residuals and reliable fitting parameters (see Table~\ref{cmds}).
Similarly for both the CMDs, the worst fitting is found for the middle RGB (within 0.1 mag in colour) and blue HB (up to 0.8 mag for the $F438W$ magnitude).
However, both DSED and BaSTI cannot explain the segregation of the RGB populations by a reliable $\Delta Y=0.025$.
Only the DSED isochrones with a very large $\Delta Y=0.08$ can do it.

Anyway, we do not use the results from UV, UV--optical, optical--IR, and IR--IR pairs (including those presented in Table~\ref{cmds}) for our 
final results, except the reddening $E(V-J_\mathrm{2MASS})$ derived for the \citetalias{hendricks2012} data set and used for our final reddening estimate.
Fig.~\ref{ngc6723law} shows that $E(V-J_\mathrm{2MASS})$ (open red circle) agrees with other results.

Another example of an issue in the UV is the $U$ photometry for NGC\,6723 (but not for NGC\,6362). Both the independent data sets with it, 
\citet{alcaino1999} and \citetalias{stetson2019}, are consitent in their unreliable $U$ magnitudes: they should be about 0.12 mag fainter. 
Otherwise, a large negative reddening $E(U-B)$ is derived in the isochrone fitting.
Thus, we reject the $U$ filter for NGC\,6723 from our consideration.

{\it HST} WFC3 $F438W$ and ACS $F606W$ filters would be a valuable proxy for the $B$ and $V$ filters. 
However, belonging to the different detectors (WFC3 and ACS), these filters may allow a small instrumental systematic error in the $F438W-F606W$ colour.
Indeed, Figs~\ref{ngc6362law} and \ref{ngc6723law} show a fracture of the {\it HST} empirical extinction laws (red diamonds) between the 
$F438W$ and $F606W$ filters. This fracture seems to be similar for BaSTI and DSEP. This forces us to exclude the WFC3 filters for our final estimates
of [Fe/H], age, distance, and reddening.

For NGC\,6362, the \citetalias{zloczewski2012}, \citetalias{narloch2017}, and \citetalias{stetson2019} data sets provide photometry in the $B$ and $V$ filters.
We cross-identify these data sets for a direct comparison and discover their magnitude differences to be up to 0.04 mag as some functions of magnitude. 
However, these functions are rather correlated for $B$ and $V$, providing only small systematic differences within 0.02 mag between their $B-V$ colours.
Such magnitude and colour differences at a level of few hundredths of a magnitude are common and expected \citepalias{stetson2019}.
Comparing the descriptions of these three data sets, we conclude that the reason for these inconsistencies is unknown. However, we lack justification to exclude any of these data sets.
Therefore, we process and fit them as is. The resulting parameters of the three data sets are similar, as seen in Table~\ref{cmds}. 
Also, the similarity of their extinctions is seen from Fig.~\ref{ngc6362law}. Therefore, we use all three data sets for our final results.

In order to see all systematic differences between the data sets in Figs~\ref{ngc6362law} and \ref{ngc6723law}, we do not adjust the data sets with similar filters, 
in contrast to \citetalias{ngc6205} and \citetalias{ngc288}.

\subsection{HB}
\label{hb}

In our CMDs, we show both the primordial and helium-enriched HB isochrones for the whole colour range covered by the HB stars in order to show that
the primordial and helium-enriched populations dominate the red and blue HB, respectively, in accordance with 
\citet{mucciarelli2016,heber2016}.

Our fitting of the observed colour distribution of the HB stars by the BaSTI extended HB set allows us to estimate the mass range for the majority 
of the HB stars (separately for the populations): 
$0.62-0.70$ and $0.58-0.61$ $M_{\sun}$ for the primordial and helium-enriched populations of NGC\,6362, respectively; 
$0.63-0.70$ and $0.55-0.61$ $M_{\sun}$ for the primordial and helium-enriched populations of NGC\,6723, respectively.
These estimates are obtained consistently for all CMDs, which are rich in HB stars.
Yet, these estimates may be uncertain due to the incompleteness of our HB samples and ambiguity regarding the population to which some stars belong.
However, these estimates can be compared with those from \citet{tailo2020}, who combine {\it HST} photometry and stellar population models to infer 
the average HB masses. 
They obtain $0.64\pm0.03$ and $0.60\pm0.03$ $M_{\sun}$ for the primordial and helium-enriched populations of NGC\,6362, respectively, 
and $0.65\pm0.02$ and $0.56\pm0.02$ $M_{\sun}$ for the primordial and helium-enriched populations of NGC\,6723, respectively.
Since we use different theoretical models from \citet{tailo2020}, a good agreement of these results is remarkable.

\begin{figure}
\includegraphics{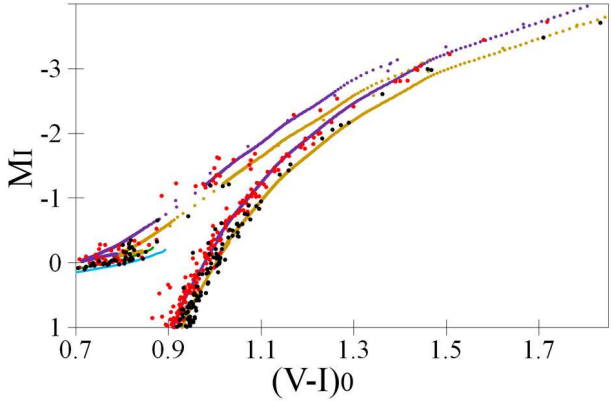}
\caption{Hertzsprung--Russell diagram $(V-I)_0$ versus $M_\mathrm{I}$ for NGC\,6362 and NGC\,6723 with the {\it Gaia} EDR3 members of the \citetalias{stetson2019} data sets:
NGC\,6362 stars (black dots) are fitted by the best fitting BaSTI isochrone for [Fe/H]$=-0.9$, $Y=0.275$ and age 12 Gyr (brown curve) and
by the best fitting BaSTI ZAHB for [Fe/H]$=-0.9$ and $Y=0.25$ (blue curve), NGC\,6723 stars (red dots) are fitted by the best fitting BaSTI isochrone for [Fe/H]$=-1.05$, 
$Y=0.275$ and age 12.5 Gyr (purple curve) and by the best fitting BaSTI ZAHB for [Fe/H]$=-1.05$ and $Y=0.25$ (green curve).
All the data are corrected for reddening, extinction, and distance estimates from Table~\ref{cmds}.
}
\label{stetson_vi_rgb}
\end{figure}

\subsection{Metallicity}
\label{metal}

An example of the RGB slope fitting is presented in Fig.~\ref{stetson_vi_rgb}.
This shows the Hertzsprung--Russell diagram for the {\it Gaia} EDR3 members of the \citetalias{stetson2019} data sets in both clusters.
All the data are corrected for reddening, extinction, and distance estimates from Table~\ref{cmds}.
The RGBs are fitted by isochrones with different [Fe/H] and ages.
Helium primordial and helium-enriched isochrones are shown for the HBs and RGBs, respectively, following our suggestions about their helium 
abundance in these clusters (although the blue HB, out of this figure, is helium-enriched).

By use of 11 independent optical CMDs with the well-populated RGB, we estimate the average [Fe/H]$=-1.10$ and $-0.97$ for NGC\,6362 by use of DSED and BaSTI, respectively.
We adopt their mean [Fe/H]$=-1.04\pm0.07$ (with uncertainty as half the difference between BaSTI and DSED) 
as our final [Fe/H] estimate for NGC\,6362 and use it for CMDs with a sparsely populated RGB.

Similarly, by use of 10 independent optical CMDs with the well-populated RGB, we estimate the average [Fe/H]$=-1.15$ and $-1.04$ for NGC\,6723
by use of DSED and BaSTI, respectively, and adopt their mean [Fe/H]$=-1.09\pm0.06$ as our final [Fe/H] estimate for NGC\,6723.

Our [Fe/H] estimates adequately agree with those from \citet{dotter2010} obtained in their fitting of the same {\it HST} ACS and WFPC2 photometry
by DSED isochrones: [Fe/H]$=-1.1$ for NGC\,6362 and [Fe/H]$=-1.0$ for NGC\,6723.

\citet{kerber2018} fit the HB, RGB, SGB, and MS of NGC\,6362 in a CMD with other {\it HST} data by use of the BaSTI and DSED isochrones.
They obtain [Fe/H]$=-1.15\pm0.08$ both for BaSTI and DSED in agreement with our results.

Our [Fe/H] estimates agree with those from \citet{valcin2020}, obtained from isochrone fitting of the {\it HST} ACS data by DSED:
[Fe/H]$=-1.11^{+0.14}_{-0.11}$ and $-1.06^{+0.07}_{-0.15}$ for NGC\,6362 and NGC\,6723, respectively.

It is important to compare our [Fe/H] estimates with those from spectroscopy. 
For NGC\,6362, there is good agreement within $\sigma$ with [Fe/H]$=-1.09\pm0.01$ from \citet{mucciarelli2016} and $-1.07\pm0.01$ from \citet{massari2017}. 
For NGC\,6723, there is good agreement within $\sigma$ with [Fe/H]$=-1.22\pm0.08$ from \citet{gratton2015} and [Fe/H]$=-0.98\pm0.08$ from \citet{rojas2016}. 
However, the estimate [Fe/H]$=-0.93\pm0.05$ from \citet{crestani2019} is less consistent with ours, i.e. at a level of $1.5\sigma$.

Cluster RR~Lyrae stars with accurately measured properties constrain cluster metallicity and some other parameters.
For RR~Lyrae stars in NGC\,6362, \citet{arellano2018} obtain [Fe/H]$=-1.066\pm0.126$, $E(B-V)=0.063\pm0.024$, and distance modulus $(m-M)_V=14.69\pm0.08$ from time-series 
photometry of the RRab stars, with similar results for the RRc stars. These results perfectly agree with ours.
Alternative consideration of the same data by \citet{vandenberg2018} provides a similar [Fe/H] within its uncertainties.
For RR~Lyrae stars in NGC\,6723 \citet{lee2014} obtain [Fe/H]$=-1.23\pm0.11$, $E(B-V)=0.061\pm0.014$, distance modulus $14.65\pm0.05$,
equivalent to a distance $8.47\pm0.17$ kpc. All these findings agree with ours.

Photometric and spectroscopic observations of detached eclipsing binaries allow one to derive the masses, radii, and luminosities of their component stars. 
The obtained mass--radius, mass--luminosity, and colour--magnitude relations can be compared with isochrone predictions to verify [Fe/H], abundance, and age of a cluster 
containing such binaries.
For two eclipsing binaries in NGC\,6362, \citet{kaluzhny2015} obtained reliable results with fitting by DSED, assuming [Fe/H]$=-1.07$. 
They verified that a change from [Fe/H]$=-1.07$ to $-1.15$ `does not influence the parameters of the binaries in any significant way'. 
However, both the components of a binary provide a good isochrone fitting only
(i) if their ages differ by $1.5\pm0.7$ Gyr \citep{kaluzhny2015} or 
(ii) if their helium abundances differ by about $\Delta Y\approx0.02$ \citep{vandenberg2018}, or 
(iii) if the binary has a higher C, N, and O abundance than in stellar models used \citep{vandenberg2022}.
These conditions are possible, since they do not contradict the known properties of NGC\,6362.

\begin{table}
\def\baselinestretch{1}\normalsize\normalsize
\caption[]{Our NGC\,6362 and NGC\,6723 age (Gyr) and distance (kpc) estimates from optical CMDs. All the uncertainties are standard deviations of one measurement.
}
\label{agedist}
\[
\begin{tabular}{lccc}
\hline
\noalign{\smallskip}
                     & DSED           & BaSTI          &  Mean value \\      
\hline
\noalign{\smallskip}
                     &                & NGC\,6362      & \\
\noalign{\smallskip}
Mean distance        & $7.76\pm0.09$  & $7.74\pm0.16$  & $7.75\pm0.13$ \\  
Mean age             & $12.00\pm0.50$ & $12.00\pm0.50$ & $12.00\pm0.49$ \\ 
\noalign{\smallskip}
                     &                & NGC\,6723      & \\
\noalign{\smallskip}
Mean distance        & $8.14\pm0.23$  & $8.15\pm0.18$  & $8.15\pm0.20$  \\
Mean age             & $12.35\pm0.34$ & $12.50\pm0.33$ & $12.43\pm0.34$ \\  
\hline
\end{tabular}
\]
\end{table}


\subsection{Distance and age}
\label{distance}

Similar to our previous papers, the derived distances and ages for the optical range (filters with their effective wavelengths within $400<\lambda_\mathrm{eff}<1000$ nm except 
{\it HST} WFC3 $F438W$) differ from those for the UV and IR ranges, both systematically and by their standard deviations.
The optical CMDs are preferable in terms of star quantity, sample completeness, photometric accuracy, and systematic accuracy of the models.
Therefore, we use the distance and age estimates in the optical range for our final estimates.

We present our age and distance estimates for NGC\,6362 and NGC\,6723 in Table~\ref{agedist}.
Their random uncertainties are calculated as standard deviations from 11 and 10 optical CMDs for NGC\,6362 and NGC\,6723, respectively.
The right column presents the mean value and standard deviation of one measurement for the combined results of the two models.
The standard deviations for the models and for the mean values are consistent, whereas the distributions of the combined results are Gaussian.
This indicates good agreement between the models in their distances and ages. 
Therefore, our final, most probable estimates for NGC\,6362 and NGC\,6723, respectively, are as follows:
\begin{enumerate}
\item age is $12.00\pm0.10\pm0.8$ and $12.43\pm0.08\pm0.8$ Gyr (statistic and systematic uncertainties), 
\item distance is $7.75\pm0.03\pm0.15$ and $8.15\pm0.04\pm0.15$ kpc,
\item distance modulus $(m-M)_0$ is $14.45\pm0.01\pm0.04$ and $14.56\pm0.01\pm0.04$ mag,
\item apparent $V$-band distance modulus $(m-M)_\mathrm{V}$ is $14.64\pm0.03\pm0.06$ and $14.80\pm0.03\pm0.06$ mag.
\end{enumerate}
Here the random uncertainties equal the standard deviations of the combined samples divided by the square root of the number of the CMDs and models used
(11 CMDs by 2 models and 10 CMDs by 2 models for NGC\,6362 and NGC\,6723, respectively).

Note that, in contrast to this study, DSED and BaSTI provided inconsistent age estimates in our previous papers.
The reason is that the BaSTI isochrones were affected by a computational issue related to input physics and atomic diffusion treatment, which was solved a few months later 
(Cassisi, private communication).
Hence, the BaSTI age estimates from our previous papers should be revisited.

Such a good agreement of the DSED and BaSTI estimates for both age and distance inspires our discussion of their systematic errors (see Sect.~\ref{systematics}), 
which can be similar in both the models and, hence, should be considered separately.
Systematic uncertainties of $\pm0.15$ kpc and $0.8$ Gyr are assigned to our distance and age estimates, respectively.
As noted in Sect.~\ref{systematics}, since we adopt these systematic uncertainties from our study-to-study comparison, {\it a posteriori} comparison of our results with those 
from the studies has little sense. Therefore, the following comparison pays less attention to the literature estimates from isochrone-to-CMD studies, while more to other methods.

Our distance estimates agree with those from the recent compilation of all distance determinations by \citet{baumgardt2021},\footnote{This compilation
is so comprehensive that our distance estimates do not need a comparison with individual estimates from the literature.}
presented in Table~\ref{properties}, within $1.1\sigma$ and $0.8\sigma$ of their stated statistical uncertainties for NGC\,6362 and NGC\,6723, respectively, and even better --
for the systematic uncertainties.

Note that both our and \citet{baumgardt2021}'s estimates for NGC\,6723 differ significantly from that of \citet{harris} (see Table~\ref{properties}).
An impact of the Corona Australis cloud complex can be the reason for this difference in the distance estimates.
Indeed, the distance estimate of \citet{harris} may be derived from the apparent magnitudes with underestimated extinction/reddening due to ignoring of the 
complex. In this case, the difference between our distance estimates and that of \citet{harris}, i.e. $8700-8150=550$ pc,
can be explained by the difference between the corresponding distance moduli $14.70-14.56=0.14$ mag and, in turn, by the underestimation of extinction by
$\Delta A_\mathrm{V}=0.14$ mag or reddening by $\Delta E(B-V)\approx0.045$ mag. 
Such an underestimation can be easily explained by ignoring the differential reddening up to $\Delta E(B-V)\approx0.13$ mag in the NGC\,6723 field (see Sect.~\ref{difred}).

\begin{table*}
\def\baselinestretch{1}\normalsize\normalsize
\caption[]{Various parallax estimates (mas) for NGC\,6362 and NGC\,6723.
}
\label{parallax}
\[
\begin{tabular}{lcc}
\hline
\noalign{\smallskip}
 Parallax            &  NGC\,6362  &  NGC\,6723 \\
\hline
\noalign{\smallskip}
\citet{shao2019}, {\it Gaia} DR2 astrometry           & $0.123\pm0.025$ & $0.119\pm0.025$  \\
\citetalias{vasiliev2021}, {\it Gaia} EDR3 astrometry & $0.132\pm0.011$ & $0.129\pm0.011$  \\
This study, {\it Gaia} EDR3 astrometry                & $0.123\pm0.011$ & $0.123\pm0.011$  \\
This study, isochrone fitting                         & $0.129\pm0.002$ & $0.123\pm0.002$  \\
\hline
\end{tabular}
\]
\end{table*}

Our distance estimates can be converted into parallaxes for their comparison with other parallax estimates in Table~\ref{parallax} (the systematic uncertainty $\pm0.15$ kpc is 
assigned to our distances). An agreement within the stated uncertainties is seen.
Anyway, the cluster distances from the {\it Gaia} parallaxes are still less accurate than those obtained from such an isochrone fitting.

Our age estimates are within a wide variety of age estimates for these GCs from the literature.
Some differences in these estimates may be due to different approaches.
For example, \citet{oliveira2020} fit the same {\it HST} ACS data with DSED and BaSTI to obtain age estimates of $13.6\pm0.5$ and $12.6\pm0.6$ Gyr 
for NGC\,6362 and NGC\,6723, respectively.  The former agrees with our estimate much worse than the latter.
The reason may be that, in contrast to our study, they fit only the SGB and bright MS. Hence, in fact, they estimate age using only the length of the SGB.
Such an estimate may be biased by large systematics.

\begin{figure*}
\includegraphics{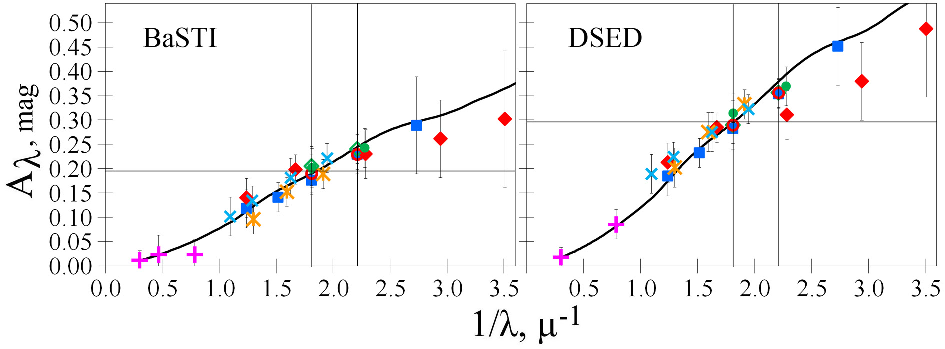}
\caption{The empirical extinction laws for NGC\,6362 from the isochrone fitting by the different models.
The data sets are:
{\it HST} ACS and WFC3 by \citet{nardiello2018} -- red diamonds;
{\it HST} WFPC2 by \citet{piotto2002} -- green circles;
{\it Gaia} -- yellow snowflakes;
\citetalias{stetson2019} -- blue squares;
SMSS -- blue inclined crosses;
\citetalias{zloczewski2012} -- open green diamonds;
\citetalias{narloch2017} -- open red circles;
IR data sets by VISTA and unWISE -- purple upright crosses.
The effective wavelengths of the $B$ and $V$ filters are denoted by the vertical lines.
The black curve shows the extinction law of \citetalias{ccm89} with $R_\mathrm{V}=3.1$ with the derived $A_\mathrm{V}$, which is shown by the horizontal line.
}
\label{ngc6362law}
\end{figure*}

\begin{figure*}
\includegraphics{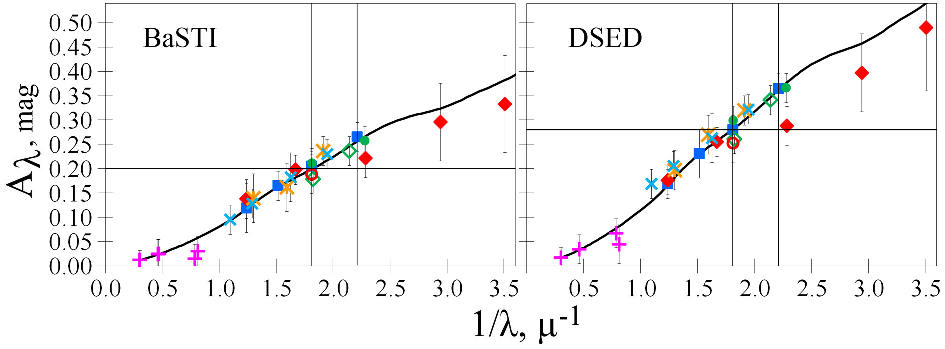}
\caption{The same as Fig.~\ref{ngc6362law} but for NGC\,6723.
The data sets are:
{\it HST} ACS and WFC3 \citet{nardiello2018} -- red diamonds;
{\it HST} WFPC2 by \citet{piotto2002} -- green circles;
{\it Gaia} -- yellow snowflakes;
\citetalias{stetson2019} -- blue squares;
SMSS -- blue inclined crosses;
\citet{lee2019} -- open green diamonds;
\citetalias{hendricks2012} ($V$ filter) -- open red circle;
IR data sets by \citetalias{hendricks2012} ($J_\mathrm{2MASS}$ and $Ks_\mathrm{2MASS}$ filters), VISTA, and unWISE -- purple upright crosses.
The effective wavelengths of the $B$ and $V$ filters are denoted by the vertical lines.
The black curve shows the extinction law of \citetalias{ccm89} with $R_\mathrm{V}=3.1$ with the derived $A_\mathrm{V}$, which is shown by the horizontal line.
}
\label{ngc6723law}
\end{figure*}

\subsection{Reddening and extinction}
\label{redext}

Reddenings, derived from CMD isochrone fitting, depend on adopted [Fe/H]. For example, \citet{kerber2018} show that a change of the adopted [Fe/H] from $-1.15$ to $-1.08$ 
for NGC\,6362 is followed by a change of the resulting $E(B-V)$ from $0.066$ to $0.053$ mag. 
[Fe/H] uncertainty is one of the largest contributors of total systematic uncertainty of derived reddening and extinction.
Its importance has been underestimated in the balance of reddening uncertainties in our previous papers.
Now we estimate the total systematic uncertainty of our reddening and extinction estimates at the level of $\sigma E(B-V)=0.02$ and 
$\sigma A_\mathrm{V}=0.06$ mag.
Consequently, for such GCs with rather low reddening and extinction, $R_\mathrm{V}$ or any other characteristic of empirical extinction law 
must be very uncertain, being a ratio of these uncertain quantities.
In addition, we take into account that both the data and models are less accurate in the UV and IR than in the optical range.
Therefore, we change our approach in deriving reddening and extinction w.r.t. our previous papers.

Specifically, we derive our reddening and extinction estimates for each pair of a data set and a model from their most informative and precise CMDs, i.e. those presenting 
the longest optical wavelength range for each data set. 
For both the clusters, these are the CMDs with the following colours:
$F606W-F814W$ for {\it HST} ACS; $F439W-F555W$ for {\it HST} WFPC2; $G_\mathrm{BP}-G_\mathrm{RP}$ for {\it Gaia} EDR3; $B-I$ for \citetalias{stetson2019};
and $g_\mathrm{SMSS}-z_\mathrm{SMSS}$ for SMSS. 
In addition, for NGC\,6362, we use $B-V$ from \citetalias{zloczewski2012} and \citetalias{narloch2017}, whereas 
for NGC\,6723, we use $V-J_\mathrm{2MASS}$ from \citetalias{hendricks2012}, and $b-y$ from \citet{lee2019}.
Thus, we use seven data sets for each cluster.
In order to combine these reddening estimates into our final estimates, we use the \citetalias{ccm89} extinction law with $R_\mathrm{V}=3.1$.

Similar to our previous papers, we verify the usage of this law by combining all the derived reddening estimates into empirical extinction laws.
Briefly, we cross-identify the data sets with the VISTA and unWISE data sets\footnote{Except the data sets of \citet{piotto2002} and \citetalias{hendricks2012}. The latter uses 
2MASS instead of VISTA or unWISE.}
and use their IR extinctions to calculate extinctions in all filters from the derived reddenings. For example,
\begin{equation}
\label{avaw1}
A_\mathrm{V}=(A_\mathrm{V}-A_\mathrm{W1})+A_\mathrm{W1}=E(V-W1)+A_\mathrm{W1}
\end{equation}
which is derived from reddening $E(V-W1)$ and a reasonable estimate of very low extinction $A_\mathrm{W1}$.

Figs~\ref{ngc6362law} and \ref{ngc6723law} show that the obtained empirical extinction laws agree with the \citetalias{ccm89} extinction 
law when $R_\mathrm{V}=3.1$.
Also, Figs~\ref{ngc6362law} and \ref{ngc6723law} act as crucial verifiers of the systematic agreement of different data sets within 
few hundredths of a magnitude.
Indeed, the data sets show a low scatter around the extinction law curve.
The only issue is a slight inconsistency between the {\it HST} ACS and WFC3 filters, providing a fracture in the {\it HST} empirical extinction law
between the $F438W$ and $F606W$ filters (see Sect.~\ref{issues}). As noted earlier, this forces us to exclude the WFC3 filters from any final estimate.

\begin{table}
\def\baselinestretch{1}\normalsize\normalsize
\caption[]{The estimates of $E(B-V)$ by use of the various models. The model estimates are mean values for seven data sets with the standard deviations of the mean values.
The final values are the averages of the models with their uncertainties as half the differences between the model estimates.
}
\label{av}
\[
\begin{tabular}{lcc}
\hline
\noalign{\smallskip}
                & NGC\,6362         & NGC\,6723       \\
\hline
\noalign{\smallskip}
BaSTI                & $0.047\pm0.004$   & $0.059\pm0.004$  \\    
DSED                 & $0.066\pm0.005$   & $0.077\pm0.006$  \\    
\noalign{\smallskip}
Final value          & $0.056\pm0.011$   & $0.068\pm0.009$ \\   
\hline
\end{tabular}
\]
\end{table}

The final reddening estimates are presented in Table~\ref{av}. It is seen that the DSED estimates are systematically higher than their BaSTI equivalents by about 0.02 mag. 
This is due to systematically lower [Fe/H] of the DSED best-fitting isochrones (see Sect.~\ref{metal}).

A systematic uncertainty of $\pm0.02$ mag is assigned to our final $E(B-V)$ estimates (see Sect.~\ref{systematics}).
This is about twice as large as half the differences between the model estimates in Table~\ref{av}.
Accordingly, the systematic uncertainty of $\pm0.06$ mag is assigned to our final extinction estimates $A_\mathrm{V}=0.19$ and $0.24$ mag for NGC\,6362 and NGC\,6723, respectively.
Here we use the ratio $A_\mathrm{V}=3.48E(B-V)$, taking into account intrinsic spectral energy distribution of rather cool and metal-poor stars of GCs (see \citealt{casagrande2014}).
It is worth noting that for NGC\,6723, these are the reddening and extinction estimates at the centre of its field.
Its differential reddening (see Sect.~\ref{difred}) should be taken into account for a proper estimate for a star or a part of the field.
Note that a similar level of the reddening precision in Table~\ref{av} for NGC\,6362 and NGC\,6723 confirms our account of differential reddening.

Our $A_\mathrm{V}$ estimates are lower than those from \citet{wagner2017}, who use a Bayesian single-population analysis for the {\it HST} ACS data and obtain 
$A_\mathrm{V}=0.248^{+0.001}_{-0.002}$ and $0.286^{+0.002}_{-0.002}$ for NGC\,6362 and NGC\,6723, respectively.
However, our $A_\mathrm{V}$ estimates are higher than those from \citet{valcin2020}, who use isochrone fitting of the {\it HST} ACS data by DSED and obtain
$A_\mathrm{V}=0.16\pm0.02$ and $0.20\pm0.03$ for NGC\,6362 and NGC\,6723, respectively.
The estimates of these studies are inconsistent each other. Yet, our large systematic uncertainty allows our estimates to be consistent with both of them.

Our $E(B-V)$ estimates agree with those from the isochrone fitting of the same or similar {\it HST} data with DSED by \citet{dotter2010}: 
$E(B-V)=0.070$ for NGC\,6362 and $0.073$ for NGC\,6723;
with DSED and BaSTI, but only using the SGB and MS, by \citet{oliveira2020}: $E(B-V)=0.04\pm0.01$ for NGC\,6362 and $0.06\pm0.01$ for NGC\,6723;
with DSED and BaSTI by \citet{kerber2018}: $E(B-V)=0.07\pm0.01$ for NGC\,6362; and
with a different model by \citet{vandenberg2013}: $E(B-V)=0.076$ for NGC\,6362 and $0.070$ for NGC\,6723.

As noted in Sect.~\ref{metal}, our reddening estimates agree with those obtained for cluster RR~Lyrae stars by \citet{arellano2018} ($E(B-V)=0.063\pm0.024$ for NGC\,6362) 
and \citet{lee2014} ($E(B-V)=0.061\pm0.014$ for NGC\,6723).

We compare our $E(B-V)$ estimates with those in Table~\ref{properties}.
For NGC\,6362, our estimate is near the lowest estimate of \citet{schlaflyfinkbeiner2011}, albeit comparable with that of \citetalias{sfd98}.
The reason for the deviation of our estimate from those of \citet{harris} and \citet{planck} is not known.
However, the inconsistency of the estimates of \citetalias{sfd98} and \citet{planck}, both calibrated from dust emission, may suggest a gradient of dust temperature or other 
inhomogeneity of the dust medium in the NGC\,6362 field.
For NGC\,6723, our estimate agrees with only that of \citet{harris}, which is apparently related to a majority of the cluster members and, hence, to the centre of the cluster. 
Naturally, our estimate is much lower than those from \citet{schlaflyfinkbeiner2011}, \citetalias{sfd98}, and \citet{planck}, which represent the whole cluster field with the 
reddening rise to its southern periphery.

\begin{table}
\def\baselinestretch{1}\normalsize\normalsize
\caption[]{The relative estimates of the derived [Fe/H] (dex), distance (kpc), age (Gyr), and $E(B-V)$ (mag) in the sense 
`NGC\,6723 minus NGC\,6362'. The mean values are shown with their uncertainties.
}
\label{relative}
\[
\begin{tabular}{lccc}
\hline
\noalign{\smallskip}
             &  DSED    &  BaSTI   &  Mean value \\
\hline
\noalign{\smallskip}
[Fe/H]               & $-0.039$ & $-0.067$ & $-0.053\pm0.014$  \\
Distance             & 0.41     & 0.48     & $0.44\pm0.04$   \\
Age                  & 0.39     & 0.61     & $0.50\pm0.11$   \\
$E(B-V)$             & 0.0070   & 0.0078   & $0.0074\pm0.0016$  \\
\hline
\end{tabular}
\]
\end{table}

\subsection{Relative estimates and second parameter}

As noted earlier, NGC\,6362 and NGC\,6723 are rich in precise photometry presented in the twin data sets.
We use 9 independent optical CMDs to derive relative (in the sense `NGC\,6723 minus NGC\,6362') estimates of [Fe/H], distance, age, and $A_\mathrm{V}$ 
from isochrone fitting to each data set separately for each model.
Other CMDs are eliminated from this procedure, since they engage the UV or IR filters. 
The only eliminated independent optical CMD, with the $F438W-F606W$ colour, has the issue discussed in Sect.~\ref{issues}.
The models are consistent in their relative estimates, which are presented in Table~\ref{relative}.

For each parameter, we calculate two kinds of uncertainty: half the difference between the BaSTI and DSED estimates and standard deviation of the mean
value calculated as standard deviation of the combined sample (both BaSTI and DSED estimates) divided by the square root of the estimates.
The latter makes sense, since the distribution of such a combined sample by each parameter is nearly Gaussian.
The largest among two uncertainties is adopted as the final uncertainty of each parameter in Table~\ref{relative}.
As expected, the relative estimates are close to the differences between the absolute estimates, but they are much more precise, as seen from Table~\ref{relative}:
NGC\,6723 is $0.44\pm0.04$ kpc further, $0.5\pm0.1$ Gyr older, $\Delta E(B-V)=0.007\pm0.002$ more reddened, and with $0.05\pm0.01$ dex lower [Fe/H] than NGC\,6362. 
These uncertainties emphasize very high sensitivity of our approach.

The lower metallicity of NGC\,6723 w.r.t. NGC\,6362 is manifested in their different RGB slopes (for example, see Fig.~\ref{stetson_vi_rgb}). 
The higher age of NGC\,6723 w.r.t. NGC\,6362 is manifested in their different SGB lengths and HB--SGB magnitude differences (for example, see Figs~\ref{cmd1}).
Such a combination of age and metallicity explains the difference in the HB morphology of these clusters mentioned in Sect.~\ref{clusters}: 
NGC\,6362 and NGC\,6723 are richer in red and blue HB stars, respectively, and NGC\,6723 has a longer blue hook of very hot HB stars, as seen in CMDs.
This HB morphology difference can be expressed as the HB types
\footnote{The HB type is defined as $(N_B-N_R)/(N_B+N_V+N_R)$, where $N_B$, $N_V$, and $N_R$ are the number of stars that lie blueward of the 
instability strip, the number of RR~Lyrae variables, and the number of stars that lie redward of the instability strip, respectively \citep{lee1994}.}
of these clusters ($-0.08$ for NGC\,6362 and $-0.58$ for NGC\,6723 from \citealt{mackey2005}), 
or as their median colour difference between the HB and RGB ($\Delta (V-I)=0.247\pm0.012$ for NGC\,6362 and $0.371\pm0.039$ for NGC\,6723 from \citealt{dotter2010}).
Both these characteristics represent a bluer HB of NGC\,6723. 

Both lower metallicity and higher age amplify the HB morphology difference, making the NGC\,6723's HB bluer. Thus, all known parameters of these clusters, including our
relative age and metallicity estimates, suggest that age is the second parameter for these clusters.

\section{Conclusions}
\label{conclusions}

This study generally follows \citetalias{ngc5904}, \citetalias{ngc6205}, and \citetalias{ngc288} in their approach to estimate some key parameters
of Galactic globular clusters by fitting model isochrones to multiband photometry. 
To verify the sensitivity of our approach, we have considered the pair NGC\,6362 and NGC\,6723 with similar metallicity, distance, age, and extinction.
In addition to distance, age, and extinction in various bands as the parameters in our previous studies, we also derived [Fe/H] of the clusters through isochrone fitting of the 
slope of the RGB in some CMDs.
The obtained metallicities, [Fe/H]$=-1.04\pm0.07$ and $-1.09\pm0.06$ for NGC\,6362 and NGC\,6723, respectively, agree with the spectroscopic estimates from the literature.

We used the photometry in 22 and 26 filters for NGC\,6362 and NGC\,6723, respectively, from the {\it HST}, {\it Gaia} EDR3, SMSS DR3, VISTA VHS DR5, unWISE, and other data sets.
These filters span a wavelength range from about 230 to 4060\,nm, i.e. from the UV to mid-IR. As in our previous studies, some data sets were cross-identified with each other.
This allowed us to
(i) estimate systematic differences of the data sets and
(ii) use the VISTA and unWISE photometry with nearly zero extinction for determination of extinction in all other filters and verification of agreement of the empirical extinction 
laws with the \citetalias{ccm89} law when $R_\mathrm{V}=3.1$.

As in \citetalias{ngc288}, to fit the data, we used the DSED and BaSTI theoretical models of stellar evolution for $\alpha$--enriched populations
with primordial and enhanced helium abundance.
The models differ in their physics and predictions for [Fe/H] and reddening, but those for age and distance are consistent. 
DSED provides [Fe/H] that is about 0.12 dex systematically lower than BaSTI and $\Delta E(B-V)\approx0.02$ mag systematically higher than BaSTI.

For NGC\,6362 and NGC\,6723, we derived the distances $7.75\pm0.03\pm0.15$ and $8.15\pm0.04\pm0.15$ kpc (statistic and systematic uncertainties), 
distance moduli $14.45\pm0.01\pm0.04$ and $14.56\pm0.01\pm0.04$ mag,
apparent $V$-band distance moduli $14.64\pm0.03\pm0.06$ and $14.80\pm0.03\pm0.06$ mag, 
ages $12.00\pm0.10\pm0.80$ and $12.43\pm0.08\pm0.80$ Gyr, 
extinctions $A_\mathrm{V}=0.19\pm0.04\pm0.06$ and $0.24\pm0.03\pm0.06$ mag, and 
reddenings $E(B-V)=0.056\pm0.01\pm0.02$ and $0.068\pm0.01\pm0.02$ mag, respectively.

The use of the twin data sets from {\it Gaia} EDR3, SMSS, \citetalias{stetson2019}, \citet{nardiello2018}, and \citet{piotto2002}, for both the clusters allowed us to obtain very 
precise relative estimates of the parameters.
We found that NGC\,6723 is $0.44\pm0.04$ kpc further, $0.5\pm0.1$ Gyr older, $\Delta E(B-V)=0.007\pm0.002$ more reddened, and with $0.05\pm0.01$ dex lower [Fe/H] than NGC\,6362.
These uncertainties show a high sensitivity of our approach. These differences in age and metallicity explain the difference of the HB morphology between these clusters. 
This suggests age as the second parameter for NGC\,6362 and NGC\,6723.

We found a strong differential reddening of about $\Delta E(B-V)\approx0.14$ mag across 17 arcminutes of the field of NGC\,6723 due to its proximity to the Corona Australis cloud complex.
This differential reddening may explain a large diversity of the reddening/extinction estimates for NGC\,6723 from the literature.
Moreover, the influence of the complex can explain the apparently wrong estimate of the NGC\,6723 distance in the data base of \citet{harris}.

Using the {\it Gaia} EDR3 data, we provided the lists of reliable members of the clusters and systemic proper motions with their total 
(systematic plus random) uncertainties in mas\,yr$^{-1}$:
$$\mu_{\alpha}\cos(\delta)=-5.512\pm0.024,\; \mu_{\delta}=-4.780\pm0.024$$
$$\mu_{\alpha}\cos(\delta)=1.021\pm0.026,\; \mu_{\delta}=-2.427\pm0.026$$
for NGC\,6362 and NGC\,6723, respectively.

\section*{Acknowledgements}

We acknowledge financial support from the Russian Science Foundation (grant no. 20--72--10052).

J.-W.L.\ acknowledges financial support from the Basic Science Research Program (grant no.\ 2019R1A2C2086290) through the National Research Foundation of Korea (NRF).

We thank the anonymous reviewers for useful comments.
We thank Charles Bonatto for discussion of differential reddening,
Eugenio Carretta for discussion of cluster metallicity,
Santi Cassisi for providing the valuable BaSTI isochrones and his useful comments,
Massimo Dall'Ora for his discussion of NGC\,6723 data sets,
Aaron Dotter for his comments on DSED,
Christopher Onken, Taisia Rahmatulina and Sergey Antonov for their help to access the SkyMapper Southern Sky Survey DR3,
Peter Stetson for providing the valuable $UBVRI$ photometry,
Don VandenBerg and Eugene Vasiliev for their useful comments.
We thank Michal Rozyczka for providing the data for NGC\,6362, which were gathered within the CASE project conducted at the Nicolaus Copernicus 
Astronomical Center of the Polish Academy of Sciences.

This study make use of data from the Cerro Tololo Inter-American Observatory 1 m telescope, which is operated by the SMARTS consortium.
This research makes use of Filtergraph \citep{filtergraph}, an online data visualization tool developed at Vanderbilt University through
the Vanderbilt Initiative in Data-intensive Astrophysics (VIDA) and the Frist Center for Autism and Innovation
(FCAI, \url{https://filtergraph.com}).
The resources of the Centre de Donn\'ees astronomiques de Strasbourg, Strasbourg, France
(\url{http://cds.u-strasbg.fr}), including the SIMBAD database, the VizieR catalogue access tool and the X-Match service, were widely used in this study.
This work has made use of BaSTI and DSED web tools.
This work makes use of data from the European Space Agency (ESA) mission {\it Gaia} (\url{https://www.cosmos.esa.int/gaia}), processed by the {\it Gaia}
Data Processing and Analysis Consortium (DPAC, \url{https://www.cosmos.esa.int/web/gaia/dpac/consortium}).
The Gaia archive website is \url{https://archives.esac.esa.int/gaia}.
This study is based on observations made with the NASA/ESA {\it Hubble Space Telescope}.
This publication makes use of data products from the {\it Wide-field Infrared Survey Explorer}, which is a joint project of the University of California, 
Los Angeles, and the Jet Propulsion Laboratory/California Institute of Technology.
This publication makes use of data products from the Pan-STARRS Surveys (PS1).
This study makes use of data products from the SkyMapper Southern Sky Survey. 
SkyMapper is owned and operated by The Australian National University's Research School of Astronomy and Astrophysics. 
The SkyMapper survey data were processed and provided by the SkyMapper Team at ANU. 
The SkyMapper node of the All-Sky Virtual Observatory (ASVO) is hosted at the National Computational Infrastructure (NCI).

\section*{Data availability}

The data underlying this article will be shared on reasonable request to the corresponding author.

\appendix

\section{Some CMDs of NGC\,6362 and NGC\,6723}
\label{addcmds}

\begin{figure}
\includegraphics{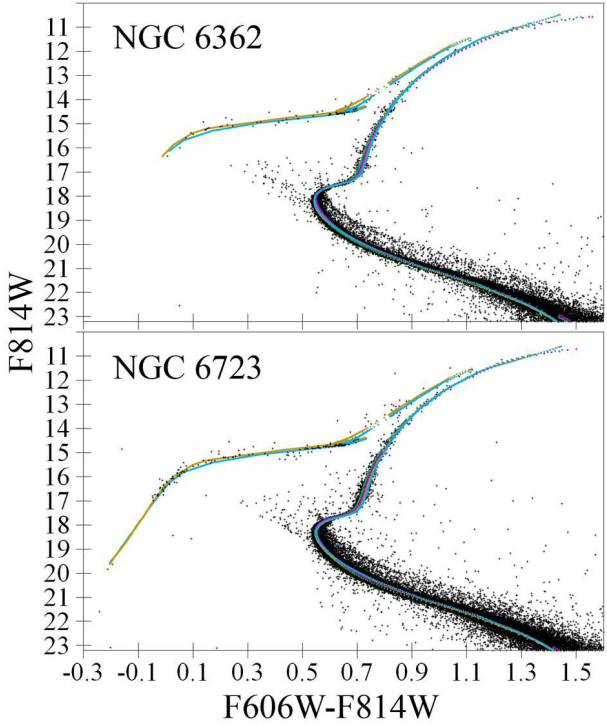}
\caption{The CMDs for NGC\,6362 and NGC\,6723 with the {\it HST} ACS $F606W-F814W$ colour. The isochrones from BaSTI for $Y\approx0.25$ (blue) and 0.275 (brown) and from 
DSED for $Y\approx0.25$ (purple) and 0.275 (green) are calculated with the best-fitting parameters from Table~\ref{cmds}.
The HB of NGC\,6723 is extended in the blue direction to show that two very hot stars may be HB members of the cluster.
}
\label{hst}
\end{figure}

\begin{figure}
\includegraphics{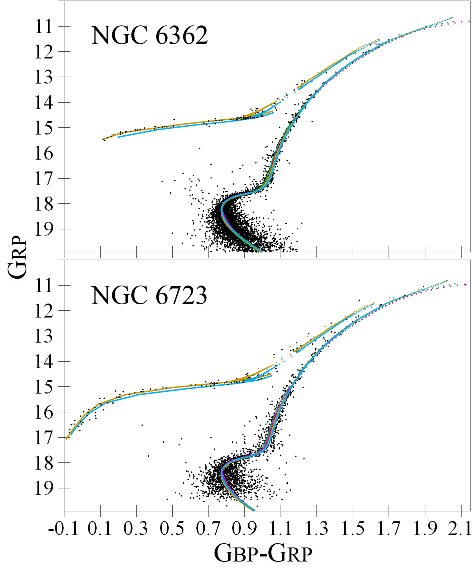}
\caption{The same as Fig.~\ref{hst} but for the {\it Gaia} EDR3 $G_\mathrm{BP}-G_\mathrm{RP}$ colour.
}
\label{gaia}
\end{figure}

\begin{figure}
\includegraphics{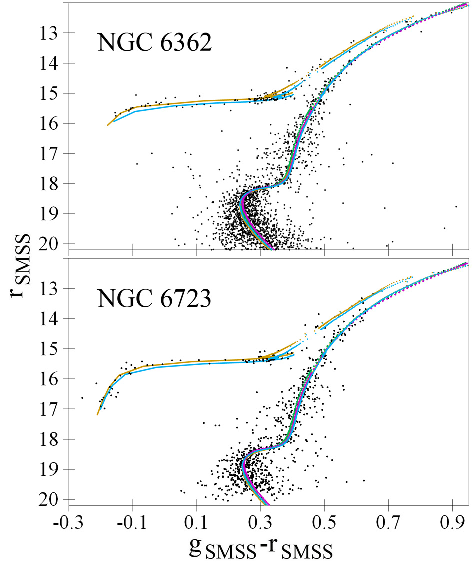}
\caption{The same as Fig.~\ref{hst} but for the SMSS $g_\mathrm{SMSS}-r_\mathrm{SMSS}$ colour.
}
\label{smss}
\end{figure}

\bsp	
\label{lastpage}
\end{document}